\documentclass[letter,11pt]{article}

\usepackage{jheppub}
\usepackage{hyperref}
\usepackage{amsmath}
\usepackage{graphicx}
\usepackage{amsmath}
\usepackage{amssymb} 
\usepackage{xspace}
\usepackage{slashed}
\usepackage{subcaption}
\usepackage{cancel}
\usepackage[normalem]{ulem}
\usepackage[dvipsnames]{xcolor}
\usepackage[utf8]{inputenc}
\usepackage[T1]{fontenc}
\usepackage{mathtools}
\usepackage{stmaryrd}
\usepackage{enumerate}
\usepackage{mathrsfs}
\usepackage[toc,page]{appendix}
\usepackage{xcolor}
\usepackage{pstricks}
\usepackage{float}
\usepackage{axodraw4j}

\newcommand{\nocontentsline}[3]{}
\newcommand{\tocless}[2]{\bgroup\let\addcontentsline=\nocontentsline#1{#2}\egroup}

\newcommand{\abs}[1]{\lvert#1\rvert}

\newcommand{\nn}{\nonumber}

\renewcommand{\max}{\mathrm{max}}

\newcommand{\bea}{\begin{eqnarray}}
	\newcommand{\eea}{\end{eqnarray}}

\def\ln{\textrm{ln}}

\def\nn{\nonumber}

\allowdisplaybreaks[2]

\DeclareUnicodeCharacter{2212}{-}

\usepackage[mathscr]{eucal}

\usepackage{marginnote}

\title{Non-universal Milan factors for QCD jets}
\author{Farid Hounat}
\affiliation{Department of Physics and Astronomy, University of Manchester, Manchester M13 9PL, United Kingdom}
\emailAdd{farid.hounat-2@postgrad.manchester.ac.uk}
\abstract{Using the dispersive method we perform a two-loop analysis of the leading non-perturbative power correction to the change in jet transverse momentum $p_T$, in the small $R$ limit of a Cambridge-Aachen jet clustering algorithm.  We frame the calculation in such a way so as to maintain connection with the universal ``Milan factor'' that corrects for the naive inclusive treatment of the leading hadronization corrections. We derive an enhancement factor that differs from the universal Milan factor computed for event-shape variables as well as the corresponding enhancement factor previously derived for the $k_t$ algorithm. Our calculation directly exploits the soft and triple-collinear limit of the QCD matrix element and phase space, which is relevant for capturing the coefficient of the leading $1/R$ power correction. As an additional check on our new calculational framework, we also independently confirm the known result for the $k_t$ algorithm.}

\begin{document}
	\maketitle
	\newpage
	
	\section{Introduction}\label{section:introduction}
	
	The LHC has collected a wide range of high quality data, and the lack of discovery of new physics beyond the standard model has shifted attention over to precision measurements and phenomenology. Events studied at the LHC are modelled as high energy perturbative interactions that evolve into a detectable final state. The initial stage of high energy collisions i.e. the hard process can be treated in perturbation theory. Remarkably accurate predictions have been made using perturbative QCD, exploiting asymptotic freedom of the underlying partonic degrees of freedom. For example, in the case of event shapes \cite{Dasgupta:2003iq} significant progress has been made in fixed-order calculations performed at next-to-next-to-leading order (NNLO), e.g. \cite{Gehrmann-DeRidder:2007nzq, Dissertori:2008cn, Gehrmann-DeRidder:2009fgd,Weinzierl:2008iv, Weinzierl:2009yz,Gehrmann:2019hwf,Alvarez:2023fhi}, and all-order resummations have been performed to next-to-next-to-leading logarithmic order (NNLL) e.g. \cite{Stewart_2011,Jouttenus_2013,Kang:2013nha,Banfi:2014sua,Becher:2015gsa, Banfi:2016zlc,Becher_2016, Tulipant:2017ybb, Banfi:2018mcq,Gao:2019ojf,Kardos:2020gty,Dasgupta:2022fim,vanBeekveld:2023lsa,Chen:2023zlx,Bhattacharya:2023qet}, and in some instances next-to-next-to-next-to-leading logarithmic order (N$^3$LL) e.g. \cite{Becher:2008cf, Chien:2010kc,Abbate:2010xh,Hoang:2014wka}. A complete description, however, requires one to account for the inevitability that the partons must form colour singlets, hadronizing into said detectable final state, resulting in calculations being complicated by hadronization corrections.
	
	Hadronization models used by Monte Carlo event generators such as the Lund hadronization model implemented in Pythia \cite{Sjostrand:2006za} and the cluster hadronization model \cite{Winter:2003tt} in Sherpa  \cite{Sherpa:2019gpd}  and Herwig \cite{Kupco:1998fx,Bahr:2008pv} are excellent phenomenological tools. It is however difficult to understand uncertainties due to the model dependency and the number of parameters available to tune. In addition, if the parton level prediction of the event generator differs from a given fixed-order or resummed analytical calculation, then adding the hadronization correction extracted from the event generator to that calculation is inconsistent. This motivates the use of analytic methods which exploit soft-gluon universality, have fewer free parameters needed to quantify hadronization effects and are more tailored to the perturbative calculation at hand.
	
	For infrared and collinear safe observables, effects of hadronization are suppressed by powers of $(\Lambda_{\text{QCD}}/Q)^n$, where $\Lambda_{\text{QCD}}$ is the energy scale at which the perturbative picture breaks down. The size of the hadronization correction depends on the observable under investigation and its sensitivity to soft-gluon radiation. Typically event shapes exhibit linear $n = 1$ power corrections. For $Q \approx M_Z$, the hadronization correction can numerically compete with terms at order $\mathcal{O}(\alpha_s^2)$ \cite{Webber, Dokshitzer_1995,Dokshitzer_1998,Dokshitzer_1998_2} and obstruct vital precision measurements such as the extraction of $\alpha_s$ \cite{Abbate:2012jh,Hoang:2015hka,Kardos:2018kqj,Nason:2023asn}. On the other hand, more inclusive observables do not exhibit such leading linear power correction (eg. cross sections, rapidity, and $p_T$ distributions of colour neutral particles \cite{Caola_2022,Beneke:1998ui,Dokshitzer_1996}). In addition to scaling with the hard scale, the coefficient of the power correction is also observable dependent \cite{Beneke:1998ui}.
	
	Power corrections are associated with renormalons, which are factorial divergences appearing in the coefficients of the perturbative expansion when one considers so called renormalon ``bubbles'' on gluon lines. One finds a factorial divergence in the perturbative expansion due to the gluon momenta in the ultraviolet (UV) and infrared (IR). One technique to address a divergent series is Borel summation. While UV renormalons are Borel resummable, there is an ambiguity in the inverse Borel transform due to IR renormalons whose size is related to non-perturbative power corrections \cite{Nason_1995,Beneke:1998ui,Beneke:2000kc,Schindler:2023cww}. 
	
	In this paper we shall focus on the topic of linear power corrections, i.e. terms of the form  $\Lambda_{\rm QCD}/Q$. As already mentioned, event shapes are an example of a class of observables that are measurable and calculable order by order in perturbation theory but are known to receive such corrections. Other examples include properties of jets, such as a jet's transverse momentum \cite{Dasgupta:2009tm,Dasgupta_2008}. A typical event begins with highly boosted partons which emit radiation and decay into a collimated spray of particles known as a jet. Jet clustering algorithms are a set of rules used to uniquely associate each particle in the event with a jet \cite{Salam_2010}. The defined jets then act as a proxy for the initiating partons, allowing one to infer important properties of the initial hard parton, such as the information about its colour factor. Jet substructure techniques such as grooming and prong finding can then be applied on top of such an algorithm to aid in tagging an underlying particle, and in turn, potentially a new resonance \cite{Dasgupta_2013,Marzani:2019hun}. Jet studies however are also complicated by hadronization corrections that can numerically compete with perturbative corrections \cite{Dasgupta_2008}, and this complication must be accounted for in order to unlock the full potential of jet substructure as a precision tool.
	
	In this article, we explore the effect different jet definitions have on power corrections to QCD observables. We build on existing calculations \cite{Dasgupta:2009tm, Dasgupta_2008} to quantify the effect hadronization has on the shift in a jet's transverse momentum $\delta p_T$, in the small $R$ limit, where $R$ is the radius of the jet. In particular we extend previous one-loop analyses of a jet's power corrections to two-loop accuracy, necessary for reasons discussed below.
	
	A one-loop analysis of power corrections can be carried out on the basis there is a correspondence between renormalons and perturbative calculations performed with an infrared cutoff \cite{Bigi:1994em,Webber:1994cp}. One introduces a small gluon mass $\lambda$ in the calculation of radiative correction diagrams \cite{Akhoury:1995sp,Dasgupta_1998}, whose effect on the phase space manifests as a $(\lambda / Q)^p$ power correction, with $p = 1$ for event shapes. In another related approach, the Dokshitzer-Webber (DW) model \cite{Dokshitzer_1995} assumes the strong coupling has an infrared-regular effective form. It makes use of a coupling whose definition is extended to the IR by setting the argument of the coupling to the transverse momentum of the soft gluon \cite{Amati:1980ch}, as is done in the perturbative case, but also assumes the Landau pole of the non-perturbative coupling is absent, rendering it finite and universal. The moment of the coupling below an IR cutoff $\mu_I$ enters as a free parameter that must be simultaneously fitted alongside $\alpha_s$.
	
	A more formal approach uses a dispersive treatment \cite{Dokshitzer_1996}, where one works with a massless gluon field and introduces a dispersive variable $\mu$, which plays the role of a small gluon mass in perturbative calculations.  The dispersive method then gives a correspondence between terms non-analytic in $\mu^2/Q^2$  and infrared renormalons. Power corrections are factorised into an observable dependent coefficient one obtains by integrating over the entire phase space except for the dispersive variable, which multiplies a universal free parameter one must fit to experiment.
	
	A deficiency in the one-loop approach lies in the fact certain observables are sensitive to additional gluon mass effects depending on how the mass of the gluon is included in definition of the observable \footnote[1]{For a concrete example, see discussion around \cite{Dokshitzer_1996} eq.(4.78)  in relation to the inclusion of the gluon mass effects in the thrust's normalisation factor used in \cite{Beneke:1995pq}.} \cite{Dasgupta_1998, Dokshitzer_1996}. In addition, Nason and Seymour pointed out that there exists a modification to power corrections at the two-loop level due to the fact one cannot naively inclusively integrate over the decay products of the non-perturbative gluon triggering the power correction when the observable in question relies on the non-inclusive kinematics of the 4-parton phase space \cite{Nason:1995np}. It has been shown that a full calculation, without such an approximation, simply results in an overall universal factor in the case of jet shape observables that depend linearly on the soft-gluon triggering the hadronization correction, known as the Milan factor \cite{Dokshitzer_1998, Dokshitzer_1998_2}. Jet clustering algorithms however depend non-linearly on the final-state parton momenta and as a result the universal Milan factor for event shapes cannot be applied. While framing the calculation as a correction to the naive approximation is not necessary, it is useful especially in our case since only the non-inclusive piece is clustering-algorithm dependent. In addition, two-loop accuracy is necessary if one is to study hadronization corrections to any observables that apply a jet clustering step, eg. soft drop \cite{Larkoski:2014wba}, as clustering algorithms are sensitive to details of the gluon branching.

	Ref. \cite{Dasgupta:2009tm} first extended existing one-loop calculations  \cite{Dasgupta_2008}  for the change in $p_T$ for small $R$ jets, to two-loop level. It calculated an enhancement factor for the $k_t$ algorithm at hadron colliders \cite{Catani:1993hr,Ellis:1993tq} that differed from the universal Milan factor known for event-shapes.

	However, the $k_t$ algorithm's tendency to cluster soft wide-angle radiation for example is not ideal at hadron colliders, as this can relate to the enhancement of underlying event and pileup contamination of the jet. Hence, in the context of jet clustering and substructure studies, the anti-$k_t$ \cite{Cacciari:2008gp} and Cambridge-Aachen (C/A) \cite{Dokshitzer:1997in,Wobisch:1998wt} algorithms have grown in popularity. The C/A algorithm is the primary choice of jet grooming algorithms such as the modified Mass Drop Tagger \cite{Dasgupta:2013ihk} and more generally soft drop \cite{Larkoski:2014wba}, due to its clustering history retaining the angular-ordered QCD dynamics which unlike the anti-$k_t$ distance measure, is related physically to parton branching dynamics. In the case of soft drop, changes in the subjet $p_T$ can result in significant changes in the groomed jet mass, and the leading hadronic correction scales as $1 / R_g$ \cite{Hoang:2019ceu, Pathak:2020iue,Pathak:2023sgi,Ferdinand:2023vaf}. These boundary effects have only been considered at an inclusive (one-loop) level, and the true non-perturbative correction to the shift in a given jet's $p_T$ using the C/A algorithm has yet to be found \cite{Dasgupta_2013}.  With this in mind we extend the analysis of \cite{Dasgupta:2009tm} to the C/A algorithm using an alternative phase-space parametrisation in which our integration variables are the angles between final state partons, and the soft and triple-collinear limit relevant to the leading power correction is taken from the start. The soft limit is responsible for linear power corrections, while the triple-collinear limit, in which all angles between final state partons are taken to be comparably small, captures the leading $1/R$ behaviour at two-loop level. We also rederive the known result for the $k_t$ algorithm  as an additional consistency check on our new phase-space parametrisation and approximations.
	\\ 
	
	The layout of this paper is as follows. In section 2 we review hadronization corrections at two-loop level, first briefly discussing the dispersive approach and then the naive one-loop approximation, as well as the corrections necessary to resolve the ambiguities in the single massive gluon calculation. In section 3, we set up the calculation of the non-inclusive correction in the context of jet clustering algorithms with an alternative phase-space parametrisation to that traditionally used in Milan factor calculations. Section 4 is dedicated to the results. Our conclusions are presented in Section 5.

	\section{Hadronization corrections at two-loop level}
	\label{section:Calculating p_T at two loop accuracy}
	In this section we will highlight the work done in \cite{Dokshitzer_1998,Dasgupta:2009tm,Dasgupta_2008}, and make connection with our calculation. Before we do that though, we will quickly review the dispersive approach to hadronization corrections.
	\subsection{The dispersive approach}
	
	The dispersive approach to dealing with hadronization corrections \cite{Dokshitzer_1996}, begins with the assumption the running coupling can be expressed as a dispersion relation
	\begin{align}\label{eq:dispersion relation}
		\alpha_s(k^2) = - \int_0^{\infty} \frac{d m^2}{m^2 + k^2}\rho(m^2),
	\end{align}
	where in the following calculations the dispersive variable, $m$, enters as the gluon mass, and $\rho(m^2)$ is the "spectral density". From this an effective coupling is defined in the following way,
	\begin{align}\label{eq:definition of effective coupling}
		\rho(m^2) = \frac{d}{d\ \ln \ m^2} \alpha_{\text{eff}}(m^2),
	\end{align}
	such that for $\alpha_s \ll 1$,
	\begin{align}
		\alpha_{\text{eff}} \approx \alpha_s.
	\end{align}

	With these definitions, the effective coupling acts as an extension of the perturbative coupling down to scales at which hadronization effects arise. Such an effective coupling can then be split into two terms,
	\begin{align}
		\alpha_{\text{eff}}(m^2) = \alpha^{\text{PT}}_{\text{eff}}(m^2) + \delta \alpha_{\text{eff}}(m^2),
	\end{align}
	where $\delta \alpha_{\text{eff}}(m^2)$ is the modification to the effective coupling with support in the non-perturbative regime, and it is moments of this modification that enter as universal non-perturbative parameters one must fit to experimental data.
	
	\subsection{$\langle \delta p_T \rangle $ to two-loop accuracy}\label{Section: delta pt}
	In this article we consider the change in jet transverse momentum $\delta p_T$ after applying a clustering step. As discussed in the introduction, the massive gluon approach is insufficient when the observable under consideration is non-inclusive with respect to the gluon decay products, and this is indeed the case for jets. It has already been shown that, in the case of a jet defined by the $k_t$ algorithm, a proper treatment results in a non-universal Milan factor \cite{Dasgupta_2008}. 
	
	Typically such calculations are split into three parts. The naive calculation, and two $\mathcal{O}\left(\alpha_s^2\right)$ corrections, one inclusive and the other non-inclusive. In the case of event shapes, this allows one to calculate an overall factor  applicable to any observable in the same universality class. While the clustering algorithms we will consider depend non-linearly on final state momenta, and hence do not have a universal Milan factor, we will show that only the non-inclusive piece is algorithm dependent and as such it allows us focus on the algorithm dependence. 
	
	Rather than introduce the three parts individually, it is useful to instead consider the full calculation to $\mathcal{O}\left(\alpha_s^2\right)$ accuracy and divide it up accordingly. The change in jet transverse momentum at  $\mathcal{O}\left(\alpha_s^2\right)$ accuracy is given by \cite{Dokshitzer_1998,Dokshitzer_1998_2,Dasgupta:2009tm}
	\begin{align}\label{eq:correct expression for delta pt}
		\langle\delta p_T\rangle =  \frac{C_F}{\pi } \int d \Gamma  \left\{\alpha_s(0) + 4 \pi \chi\right\}\delta p_T(k) + 4 C_F \int d \Gamma_2 \  \left(\frac{\alpha_s\left(m^2\right)}{4 \pi}\right)^2  \ \frac{M^2}{2!} \ \delta p_T(k_1,k_2),
	\end{align}
	where the gluon has been dressed with renormalon bubbles, which sets the argument of the coupling to the mass of the gluon, hence the first term on the right hand side featuring $\alpha_s(0)$ in the single real emission case. The second term involving $\chi$ is the virtual correction to the single emission case. The third and final term takes into account the parent gluon decaying into two quarks or gluons. Note that in the single emission case, the observable depends on the parent gluon $k$, meanwhile when integrating over the two-parton phase space it depends on the decay products $k_1, k_2$.
	
	In order to make manifest the naive, inclusive and non-inclusive piece we must first divide the final term. Let
	\begin{align}\label{eq:adding and subtracting inclusive definition}
		\delta p_T(k_1,k_2) &= \delta p_T(k_1,k_2) - \delta p_T(k_1 + k_2)  + \delta p_T(k_1 + k_2),\\
		\nn
		&= \delta {p_T}_{\rm ni} + \delta {p_T}_{\rm i},
	\end{align}
	where we have defined the inclusive definition of the observable
	\begin{equation}\label{eq:inclusive definition of the observable}
		\delta {p_T}_{\rm i} \equiv \delta p_T(k_1 + k_2),
	\end{equation}
	and the non-inclusive definition
	\begin{equation}\label{eq:non-inclusive definition of the observable}
		\delta {p_T}_{\rm ni} \equiv \delta p_T(k_1,k_2) - \delta p_T(k_1 + k_2).
	\end{equation}
	
	Note here that by construction we have introduced the term $\delta p_T(k_1 + k_2)$, in which the observable is written in terms of the reconstructed parent $k = k_1 + k_2$. One can choose $k$ to be massive or massless. While the inclusive and non-inclusive terms individually depend on this choice, the ambiguity will cancel due to how we have defined eq.\eqref{eq:adding and subtracting inclusive definition}. Traditionally in Milan factor calculations a massless reconstructed parent is used. In our evaluation of the non-inclusive correction we will use a massive definition throughout, and add a correction term at the end to make connection with the usual Milan factor. We shall return to this point in Section \ref{section:Massive gluon correction}.
	
	Integrating the inclusive piece over the two parton phase space, one can show \cite{Dokshitzer_1998_2},
	\begin{multline}\label{eq:order alpha_s^2 inclusive piece integrated over phi and z}
		\int d\Gamma_2 \left(\frac{\alpha_s\left(m^2\right)}{4\pi}\right)^2 \frac{M^2}{2!} 
		\delta {p_T}_{\rm i} = 
		\int \frac{d\alpha}{\alpha} \frac{dm^2}{m^2(m^2 + k_t^2)} dk_t^2 
		\left(\frac{\alpha_s\left(m^2\right)}{4\pi}\right)^2 \\
		\times \left(-\beta_0 + 2C_A \ln \frac{k_t^2(k_t^2 + m^2)}{m^2} \right) 
		\delta {p_T}_{\rm i},
	\end{multline}
	where in these variables, $\alpha$ is the Sudakov longitudinal energy fraction the parent gluon takes from the quark which is initiating the jet, $k_t$ is the transverse momentum of that gluon and
	\begin{equation}
		\beta_0 = \frac{11}{3} N_c - \frac{2}{3}n_f.
	\end{equation}
	
	In what follows, we shall swiftly review the naive case and inclusive correction using the traditional phase space parametrisation and massive definition of the parent, and only introduce the alternative parametrisation for the non-inclusive case we are interested in.
	\subsection{The naive case}\label{section:The naive case}
	Assuming we have a valid dispersion relation eq.\eqref{eq:dispersion relation}, at one-loop level we have
	\begin{align}\label{eq:running coupling}
		-\beta_0\left(\frac{\alpha_s}{4 \pi}\right)^2 	= \frac{d}{d \ln  \ m^2}\frac{{\alpha_{\text{eff}}(m^2)}}{4 \pi},
	\end{align}
	we can then combine the first term in eq.\eqref{eq:order alpha_s^2 inclusive piece integrated over phi and z} with the single real gluon emission case (first term in eq.\eqref{eq:correct expression for delta pt}) \cite{Dokshitzer_1998_2}
	\begin{equation}
		\langle\delta p_T\rangle^0 = 	4 C_F \int  \frac{d \alpha}{\alpha}  \frac{dm^2 dk_t^2}{k_t^2 + m^2}\left\{\frac{\alpha_{\text{eff} }(0)}{4 \pi}\delta p_t(k) - \frac{1}{m^2}\frac{d}{d \ln  \ m^2}\frac{{\alpha_{\text{eff}}(m^2)}}{4 \pi}\delta p_t(k_1 + k_2)\right\}.
	\end{equation}
	
	The observable integrated over the energy fraction $\alpha$ is defined as the \emph{trigger function} (see \cite{Dokshitzer_1998, Dokshitzer_1998_2, Dasgupta:2009tm} for details)
	\begin{equation}\label{eq:trigger function naive case}
		\Omega_0\left(k_t^2 + m^2\right) = \int_{(k_t^2 + m^2)/s}^{1} \frac{d \alpha}{\alpha} \delta p_t(k = k_1 + k_2),
	\end{equation}
	allowing us to write the above equation as follows
	\begin{align}\label{eq:naive case}
		\langle\delta p_T\rangle^0 &= 	4 C_F \int    \frac{dm^2 dk_t^2}{k_t^2 + m^2}\left\{\frac{\alpha_{\text{eff} }(0)}{4 \pi}\delta(m^2) - \frac{d}{d   m^2}\frac{{\alpha_{\text{eff}}(m^2)}}{4 \pi}\right\}\Omega_0\left(k_t^2 + m^2\right), \\
		\nn
		&= \frac{C_F}{\pi} \int    \frac{dm^2}{m^2}\alpha_{\text{eff} }(m^2)\Omega_0(m^2),
	\end{align}
	where we have extended the integration limits of $m^2$ to infinity and integrated the term involving $\frac{d}{d   m^2}\frac{{\alpha_{\text{eff}}(m^2)}}{4 \pi}$ by parts. We follow the treatment of \cite{Dasgupta_2008}, where an expression for $\delta p_T$ was derived in the context of hadron collider dijets at threshold \footnote[2]{Although we have illustrated our arguments on hadron collider jets the conclusion we shall reach for $1/R$ contributions shall also apply to $e^+e^-$ jets, due to their process independent collinear origin.}. The change in jet transverse momentum with respect to the beam direction due to the emission of a parton $k$ can be split into two terms, the first contribution when the parton recombines with the parton that emitted it, giving a mass to the "trigger jet"
	\begin{align*}
		\delta p_T^+(k_i) = -\frac{M_j^2}{2 \sqrt{s}},
	\end{align*}
	where $\sqrt{s}$ is the centre of mass energy of the collision, and a second contribution when the parton does not recombine, in which case the recoil jet receives a mass
	\begin{align*}
		\delta p_T^-(k_i) = -\frac{M_r^2}{2 \sqrt{s}}.
	\end{align*}

	Consider the Sudakov decomposition of an emission $k_i$,
	\begin{equation}
		k_i = \alpha_i p + \beta_i \bar{p} + {k_t}_i,
	\end{equation}
	where we have decomposed the vector into light-cone (Sudakov) variables. Ignoring recoil, $p$ and $\bar{p}$  are identified with the quark and anti-quark 4-vectors respectively. For partons that recombine with the emitter, the jet mass is given by
	\begin{align*}
		M_j^2 = (\sum_i k_i + p)^2  = \sum_i 2 k_i \cdot p = s \sum_i \beta_i ,
	\end{align*}
	where we have used the fact by definition $2 p \cdot \bar{p} = s$. On the other hand for partons that do not recombine, the mass of the recoil jet is given by
	\begin{align*}
		M_r^2 = (\sum_i k_i+ \bar{p})^2  = \sum_i 2 k_i \cdot \bar{p} = s \sum_i \alpha_i.
	\end{align*}
	
	In the collinear limit we are interested in, $\beta_i$ and in turn the trigger jet mass vanishes resulting in a 
	negligible contribution when the gluon $k_i$ is recombined with the jet. As a result, we only consider the case where the gluon lies outside the jet. In the naive case we only have a single emission $k$, and so 
	
	\begin{equation}
		\delta p_t(k) = p_T \alpha   \Theta_{\text{out}}(k) = p_T \alpha   \Theta(\theta_{kj} - R),
	\end{equation} 
	where $\theta_{kj}$ is the angle between emission $k$ and the jet, and $R$ is the jet radius. We have also used the fact that at threshold $p_T = \sqrt{s}/2$.
	
	We can get an expression for $\theta_{kj}$ by squaring the 4-momentum of the massive parent gluon,
	\begin{equation*}
		k = \alpha p + \beta \bar{p} + k_t,
	\end{equation*}
	squaring the above,
	\begin{equation*}
		k^2 = m^2 = \frac{p_T^2}{4} \alpha \beta -  k_t^2,
	\end{equation*}
	and rearranging,
	\begin{equation}\label{eq: relating parent angle to k_t}
		\frac{4}{p_T^2}\frac{k_t^2 + m^2 }{\alpha} = \frac{\beta}{\alpha} = \frac{k \cdot p }{ k \cdot \bar{p}} =  \frac{\theta_{kj}^2}{4},
	\end{equation} 
	where in the final line we have approximated $k$ to be massless, and taken the collinear limit. This gives us an expression for $\theta_{kj}$ and $\alpha$ allowing us to integrate the trigger function \eqref{eq:trigger function naive case}. Hence \eqref{eq:naive case} can be written as
	\begin{equation}
		\langle\delta p_T\rangle^0 = - \frac{1}{R}	\frac{C_F}{\pi} \int    \frac{dm^2}{m^2}\alpha_{\text{eff} }(m^2) m.
	\end{equation}
	
	Finally, we substitute in $\delta \alpha_{\rm eff}$, and our final answer for the naive case is given by
	\begin{equation}\label{eq:naive answer}
		\langle\delta p_T\rangle^0 = - \frac{2 A_1}{R},
	\end{equation}
	where we define non-perturbative moment of the coupling $A_1$,
	\begin{equation}\label{eq:npt moment of the coupling}
		A_1 \equiv \frac{C_F}{2 \pi}\int    \frac{dm^2}{m^2}\delta \alpha_{\text{eff} }(m^2) m.
	\end{equation}
	\subsection{The inclusive correction}\label{section: the inclusive correction}
	The inclusive correction arises from an incomplete cancellation between the virtual correction to the single gluon emission case, namely the second term in eq.\eqref{eq:correct expression for delta pt} and the second term in \eqref{eq:order alpha_s^2 inclusive piece integrated over phi and z}. While the integrands are equal, the observables definitions differ, and we are left with \cite{Dokshitzer_1998, Dokshitzer_1998_2}
	\begin{equation}
		\langle\delta p_T\rangle^{\text{inc}} = {8 C_F C_A}{\beta_0} \int dm^2 \frac{\alpha_{\rm eff}(m^2)}{4 \pi}\frac{d}{d m^2} \int \frac{d k_t^2}{k_t^2 + m^2} \ln \frac{k_t^2(k_t^2 + m^2)}{m^4}\left\{\Omega_0(k_t^2 + m^2) - \Omega_0(k_t^2)\right\},
	\end{equation}
	where
	\begin{equation}
		\Omega_0(k_t^2 + m^2) - \Omega_0(k_t^2) = -\frac{1}{R}\left(\sqrt{k_t^2 + m^2} - k_t\right).
	\end{equation}
	
	Using the convergence on the $k_t$ integration we can extend the upper limit to infinity, performing the integration over $k_t$ we arrive at
	\begin{align}
		\langle\delta p_T\rangle^{\text{inc}} &= {8 C_F C_A}{\beta_0} \int dm^2 \frac{\alpha_{\rm eff}(m^2)}{4 \pi}\frac{d}{d m^2}\left\{2 m \left( 3.299 \right)\right\}\\
		\nn
		&= 3.299 \frac{2 C_F C_A}{\pi \beta_0} \int \frac{d m^2}{m^2} m \delta \alpha_{\rm eff},
	\end{align}
	which is related to eq.\eqref{eq:naive answer} in the following way
	\begin{equation}\label{eq:inclusive correction}
		\langle\delta p_T\rangle^{\text{inc}}  = \langle\delta p_T\rangle^{0} {3.299 C_A }/{\beta_0} = \langle\delta p_T\rangle^{0} r^{(i)} .
	\end{equation}
	
	The Milan factor is to be applied to the naive case, and can be written as
	\begin{equation}\label{eq:milan factor definition}
		\mathcal{M} = 1 + r^{(i)} + r^{(n)},
	\end{equation}
	where the $1$ simply accounts the naive case and $r^{(i)}$ is the inclusive correction we have just calculated above. The rest of this paper is dedicated to calculating $r^{(n)}$ for different jet clustering algorithms.

	\subsection{Non-inclusive correction}\label{subsection: non-inclusive correction}
	
	\begin{figure}[h]
		\centering
		\includegraphics[width=0.7 \textwidth]{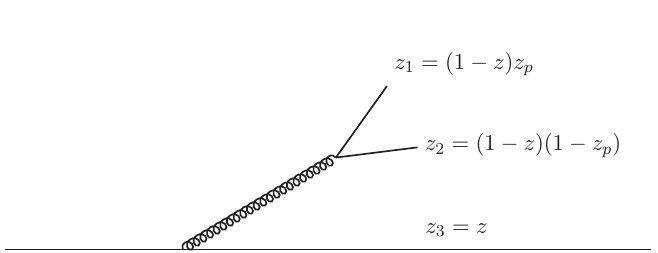}
		\caption{A virtual gluon $k$ decaying into $k_1$ and $k_2$. $k$ takes a fraction of energy $1-z$ from the quark line, and the decay products each take a fraction of energy from the parent $z_1$ and $z_2$. $z_3$ is the fraction of energy that the quark initiating the jet retains. In the soft limit, $z \rightarrow 1$.}
		\label{fig:parton config}
	\end{figure}
	The non-inclusive correction to the change in jet transverse momentum to two-loop accuracy is given by
	\begin{align}\label{eq:non inclusive correction}
		\langle\delta p_T\rangle_{\text{ni}} = 4 C_F \int d \Gamma_2 \  \left(\frac{\alpha_s}{4 \pi}\right)^2  \ \frac{M^2}{2!}  \ \delta {p_T}_{\text{ni}},
	\end{align}
	where $d\Gamma_2$ is the two-body decay phase space, and $M^2$ the corresponding squared matrix element for the "parent" gluon decaying into two partons.
	
	One extracts the non perturbative correction by first carrying out the dispersive procedure, with the dispersive variable being the parent gluon ($k$) "mass" $m$, which eventually decays into two quarks or gluons ($k_1, k_2$). $z$ is the fraction of energy the energetic quark initiating the jet retains after emitting $k$. We opt to use the following mass variable
	\begin{align}\label{eq:rho definition}
		\rho \equiv \frac{m^2}{R^2 {p_T}^2} = z_p (1-z_p)(1-z)^2 \frac{\theta_{12}^2}{R^2},
	\end{align}

	where $z_p$ denotes the fraction of energy $k_1$ takes from $k$, and similarly $(1-z_p)$ for $k_2$ (see Fig. \ref{fig:parton config}). We also define $v \equiv \frac{\theta_{12}^2}{R^2}$. In these variables eq. \eqref{eq:running coupling} is simply
	\begin{align}\label{eq:running coupling rho}
		-\beta_0 \left(\frac{\alpha_s}{4 \pi}\right)^2 	= \frac{d}{d \ln  \ \rho}\frac{{\alpha_{\text{eff}}}}{4 \pi}.
	\end{align}

	Fixing $\rho$ then substituting eq.\eqref{eq:running coupling rho} and integrating by parts as usual, we get
	\begin{align}\label{eq:non inclusive correction rho fixed}
		\langle\delta p_T\rangle_{\text{ni}} = \frac{C_F}{\pi \beta_0} \int    d \rho\   \alpha_{\text{eff}}  \  \frac{d}{d \rho}\left\{ \int  d \Gamma_2 \  \rho \ \frac{M^2}{2!} \ \delta {p_T}_{\text{ni}}  \ \delta(\rho - z_p (1-z_p)(1-z)^2 v)\right\}.
	\end{align}

	As mentioned in the introduction, we are looking for the leading $R^{-1}$ behaviour. We expect
	\begin{align}\label{eq:leading R bevaiour}
		\langle\delta p_T\rangle_{\text{ni}} \propto \frac{1}{R}\int d m \  \alpha_{\text{eff}} = \frac{p_T}{2} \int \frac{d \rho}{ \sqrt{\rho}} \ \alpha_{\text{eff}},
	\end{align}
	which suggests the term inside the curly brackets of eq.\eqref{eq:non inclusive correction rho fixed} goes as
	\begin{align}\label{eq:two body phase space integration}
		\int  d \Gamma_2 \  \rho \ \frac{M^2}{2!} \ \delta {p_T}_{\text{ni}}  \ \delta(\rho - z_p (1-z_p)(1-z)^2 v) \propto \sqrt{\rho},
	\end{align}
	and upon evaluating the exact form of \eqref{eq:two body phase space integration} and replacing the effective coupling with the non-perturbative modification $\delta \alpha_{\text{eff}}$ we get the hadronic correction
	\begin{align}\label{eq:NPT correction}
		\langle\delta p_T\rangle_{\text{ni}} = \frac{C_F}{\pi} p_T \ {r^{(n)}} \int \frac{d \rho}{\sqrt{\rho}}  \ \delta \alpha_{\text{eff}}, 
	\end{align}
	where the non-inclusive contribution to the Milan factor $r^{(n)}$ commonly defined in literature is given by
	\begin{align}\label{eq:non inclusive contribution to milan factor}
		r^{(n)} = \frac{\sqrt{\rho}}{ p_T 2 \beta_0}  \ \int  d \Gamma_2 \ \frac{M^2}{2!} \ \delta {p_T}_{\text{ni}}  \ \delta(\rho - z_p (1-z_p)(1-z)^2 v),
	\end{align}
	and in our variables the moment of the coupling $A_1$ is given by
	\begin{equation}\label{eq:npt moment of coupling in new variables}
		A_1 \equiv \frac{C_F}{2 \pi} R p_T \int \frac{d \rho}{\sqrt{\rho}}  \delta \alpha_{\text{eff}}.
	\end{equation}

	The calculation we perform amounts to evaluating eq.(\ref{eq:non inclusive contribution to milan factor}) with the corresponding clustering algorithm dependent definition of $\delta {p_T}_{\text{ni}}$.
	\section{The non-inclusive case in the new parametrisation}
	In this section we will setup the calculation of the clustering-dependent non-inclusive piece in our alternate phase space parametrisation for a general jet clustering algorithm.
	\subsection{The collinear splitting functions}
	We are considering the splittings $q  \rightarrow q q \bar{q} $, and $ q \rightarrow q   g g$. As we will see in Section \ref{section:observable}, the regions of phase space in which we receive a non-inclusive correction are around the jet boundary. This is because by definition the non-inclusive correction accounts for the mismatch between defining the observable with a massive parent gluon versus taking into account the exact kinematics of the decay products, eg. one decay product inside the jet, one outside the jet, while replacing both decay products by a single entity, the parent gluon, only allows for configurations where both partons are inside or outside the jet. In addition, we are interested in the collinear limit as that is where the leading  $1/R$  power correction comes from. As a consequence, in the small $R$ limit, the phase space of interest is the region very collinear to the initiating parton. A natural starting point is then to use the triple-collinear splitting functions, in which the angles between all partons are small and comparable. For the case where the parent gluon decays into a quark-anti-quark pair we have \cite{Catani_1999}
	\begin{align}\label{eq:qqbar collinear splitting}
		\langle \hat{P}_{\bar{q}'_1 q'_2 q_3}\rangle  = \frac{1}{2} \frac{1}{s_{123} s_{12}} \left[-\frac{t^2_{123}}{s_{12}s_{123}} + \frac{4z_3 + (z_1-z_2)^2}{z_1 + z_2} + z_1 + z_2 -\frac{s_{12}}{s_{123}}\right],
	\end{align}
	\begin{align}\label{eq:phase space variables}
		\nn s_{ij} &\equiv \frac{p_T^2}{4}z_iz_j\theta_{ij},\\
		\nn
		z_1 &= (1-z)z_p,\\
		z_2 &= (1-z)(1-z_p),\\
		\nn
		z_3 &= z,\\
		\nn
		s_{123} &\equiv \sum_{i,j} s_{ij},
	\end{align}
	\begin{align*}
		t_{123} &\equiv \frac{2}{z_1+z_2}(z_1 s_{23} - z_2 s_{13}) + \frac{(z_1 - z_2)}{(z_1+z_2)} s_{12}.
	\end{align*}

	The ${z_i}$ denote the fractions of energy of the partons involved as outlined in Section \ref{subsection: non-inclusive correction}. The angles $\theta_{ij}$ are those between the final state partons, with $i=3$ denoting the hard parton initiating the jet.
	
	Eq.\eqref{eq:qqbar collinear splitting} is accompanied by the phase space factors
	\begin{align}\label{eq:triple collinear phase space factors}
		J_1 &= \frac{1}{\pi}z_p (1-z_p) (1-z)^3 \Delta^{-\frac{1}{2}},\\
		\nn
		\Delta &= 4\theta^2_{23}\theta^2_{13} - (\theta^2_{12} - \theta^2_{13} - \theta^2_{23})^2.
	\end{align}

	Note the positivity of the Gram Determinant, $\Delta$, results in an additional constraint on the phase space. 
	As we are interested in the soft limit,  we take $z \rightarrow 1$ and eq.\eqref{eq:qqbar collinear splitting} can be written as
	
	\begin{align}\label{eq:qqbar collinear splitting in soft limit}
		\langle \hat{P}_{\bar{q}'_1 q'_2 q_3}\rangle_{\mathcal{H}_q}  &= \frac{2}{R^4}\frac{ (x-y)^2 z_p(1-z_p) - v\left(x z_p +y(1-z_p)\right)}{v^2 (1-z)^4 z_p(1-z_p) (x z_p + y (1-z_p))^2},
	\end{align}
	\begin{equation}\label{eq:defining x y and v}
		x \equiv \frac{\theta_{13}^2}{R^2},	y \equiv \frac{\theta_{23}^2}{R^2},	v \equiv \frac{\theta_{12}^2}{R^2}.
	\end{equation}
	
	The $\frac{1}{R^4}$ factor is cancelled by a factor of $R^4$ in the phase space when switching over to our new integration variables $x, y$ and $v$. The $(1-z)^4$ divergence from the matrix element combines with the $(1-z)^3$ from the phase space to give a $(1-z)$ soft divergence. The observable $\delta p_T$ vanishes in the soft limit as we are working with a soft gluon, and as a result comes with a factor of $(1-z)$ which eventually cancels the soft divergence. 
	
	The non-abelian contribution can be split into two parts. One captures the leading infrared contribution of the non-abelian piece, and the other is the less singular remainder,
	\begin{equation}\label{eq:leading infrared contribution in collinear limit}
		\langle \hat{P}_{g_1 g_2 q_3}\rangle_{\mathcal{S}} = \frac{2}{R^4}\frac{(x + y - v)}{v x y (1-z)^4 z_p^2 (1-z_p)^2},
	\end{equation}
	\begin{equation}\label{eq:finite remainder of non-abelian contribution in the soft limit}
		\langle \hat{P}_{g_1 g_2 q_3}\rangle_{\mathcal{H}_g} = \frac{1}{R^4}\frac{v (x^2 + y^2 + 6 x y - v(x+y))(xz_p + y (1-z_p)) - 2 x y (x-y)^2 z_p(1-z_p) }{v^2 x y (1-z)^4 z_p (1-z_p) (x z_p + y (1-z_p))^2 },
	\end{equation}
	where we have used the $\mathcal{S}$,$\mathcal{H}_g$ and $\mathcal{H}_q$ subscripts to make connection with \cite{Dokshitzer_1998} (see eq.\eqref{eq:define S H_g and H_q}).

	\subsection{The observable in the non-inclusive case for a general clustering algorithm}
	\label{section:observable}
	
	\begin{figure}[h]
		\centering
		\includegraphics[width=1 \textwidth]{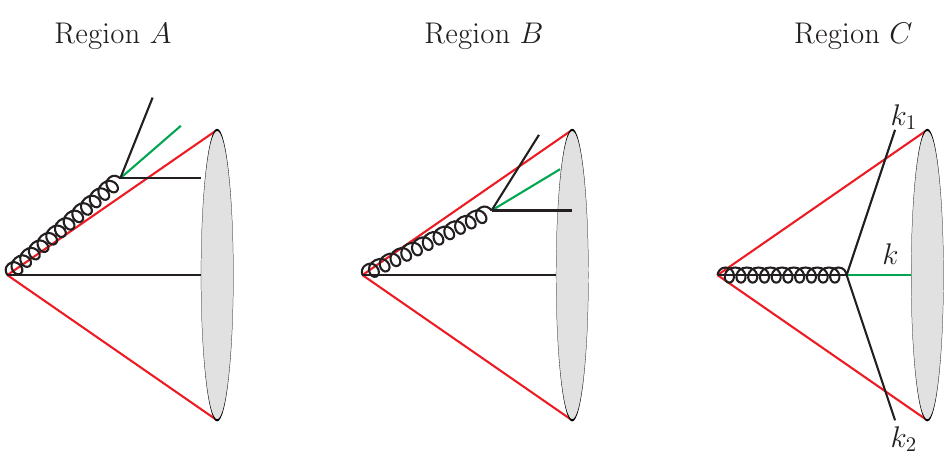}
		\caption{Three regions of phase space we must consider. The red line represents the jet boundary of radius $R$. The gluon decays into two partons (black). We receive a contribution when the reconstructed parent (green) pulls a decay product into or outside of the jet. For example, in region A $k_2$ is pulled out of the jet by clustering.}
		\label{fig:Regions of clustering}
	\end{figure}
	In this section we will derive the form of eq.\eqref{eq:non-inclusive definition of the observable} for a general clustering algorithm. The non-inclusive correction to the observable is given by the difference between the correct and naive inclusive definitions of the change in jet transverse momentum with respect to the beam direction after applying the chosen clustering algorithm. In analogy with eq.\eqref{eq:trigger function naive case}, the non-inclusive trigger function calculated in \cite{Dokshitzer_1998,Dasgupta:2009tm} is obtained after integrating eq.\eqref{eq:non-inclusive definition of the observable} over the energy fraction of the parent gluon and written in terms of dimensionless fractions of the rescaled transverse momentum. As shown in Appendix \ref{Cambridge Aachen old method}, the clustering condition manifests itself as a complicated division of the phase space. The division is shown visually in Figure 1 Appendix A of \cite{Dasgupta:2009tm}.
	
	If one retains the angular variables as in our approach, there is an intuitive way to divide the phase space. 
	We only have to consider the three following configurations which depend on locations of $k, k_1$ and $k_2$ (Fig. \ref{fig:Regions of clustering}):
	\begin{itemize}\label{configurations}
		\item $A$. $k$ out, $k_1$ in, $k_2$ out,
		\item $B$. $k$ in, $k_1$ in, $k_2$ out,
		\item $C$. $k$ in, $k_1$ out, $k_2$ out.
	\end{itemize}
	
	Regions $A$ and $B$ must be accompanied by a factor of 2 in order to account for the possibility of interchanging $k_1$ and $k_2$. 
	
	It is helpful to first divide the observable for a given clustering algorithm into two terms. The first term corresponds to the case of a rigid cone algorithm, in which one only clusters partons within a radius $R$ of the initiating hard parton to the jet, while the second term singles out the dependence on a chosen clustering algorithm,
	\begin{align}\label{eq: splitting non inclusive piece into cone + clustering}
		{\delta p_T}_{\text{ni}}&= {\delta p_T}_{\text{cone}} + {\delta p_T}_{\text{cluster}}.
	\end{align}
	
	Let us first consider the rigid cone case. $k_1$ and $k_2$  contribute independently to the correct definition  of the observable (the first term in \eqref{eq:non-inclusive definition of the observable}) depending on whether or not the partons are inside or outside the jet  \cite{Dasgupta:2009tm,Dasgupta_2008},
	\begin{align}
		{\delta p_T}(k_1, k_2) &= \delta p_T(k_1) + \delta p_T(k_2),
	\end{align}
	\begin{align}
		\delta p_T(k_i) &= \delta p_T^-(k_i)\Theta_{\text{out}}(k_i),
	\end{align}
	\begin{align}
		\delta p_T^-(k_i) = -p_T \alpha_i,
	\end{align}
	where $\alpha_i$ is  the longitudinal momentum fraction as described in Section \ref{section:The naive case}. In the inclusive case (the second term in \eqref{eq:non-inclusive definition of the observable}), the observable only depends on $k = k_1 + k_2$,
	\begin{align*}
		\delta p_T(k = k_1 + k_2) &= \delta p_T^-(k)\Theta_{\text{out}}(k).
	\end{align*}

	The observable then in the rigid cone case is then given by
	\begin{align*}
		{\delta p_T}_{\text{cone}} &= \delta p_T^-(k_1)\Theta_{\text{out}}(k_1) + \delta p_T^-(k_2)\Theta_{\text{out}}(k_2) - \delta p_T^-(k)\Theta_{\text{out}}(k),\\
		\nn
		&= -p_T z_1 \Theta_{\text{out}}(k_1) -p_T z_2\Theta_{\text{out}}(k_2) +  p_T(z_1 + z_2)\Theta_{\text{out}}(k),
	\end{align*}
	where we have used the fact $\alpha_i = z_i$. We define $\theta_p$ to be the angle between the reconstructed parent $k$ and the jet, rescaled by the jet radius $R$
	\begin{align}\label{eq:definition of theta_p}
		\theta_p \equiv \frac{\theta_{kj}}{R}.
	\end{align}

	For each of the regions we outlined above, we can write
	\begin{itemize}
		\item $A$. ${\delta p_T}_{\text{cone}, A}  = p_T  z_1 \Theta(x - 1)\Theta(1 - y)\Theta(\theta_p  - 1)$,
		\item $B$. ${\delta p_T}_{\text{cone}, B}  = - p_T z_2\Theta(x - 1)\Theta(1 - y)\Theta(1 - \theta_p)$,
		\item $C$. ${\delta p_T}_{\text{cone}, C}  = - p_T  (z_1 + z_2)\Theta(x - 1)\Theta(1 - y)\Theta(1 - \theta_p)$.
	\end{itemize}

	Now we must calculate the clustering algorithm dependence. This is a simple extension of what we have done above. Consider region $A$ where $k$ is outside the jet. In the region that the C/A algorithm combines the decay products $k_1, k_2$ into $k$, the contribution due to $k_2$ is unchanged as it was already outside the jet. However, now $k_1$ gets pulled outside the jet. So, in region $A$ the clustering correction to the observable reads
	\begin{align*}
		{\delta p_T}_{\text{cluster}, A}  = - p_T z_1 \Xi_{\rm A} \Theta(x - 1)\Theta(1 - y)\Theta(\theta_p  - 1) = - {\delta p_T}_{\text{cone}} \Xi_{\rm A},
	\end{align*}
	where we have introduced $\Xi_{\rm A}$ to encapsulate the clustering condition for a non-zero contribution to region $A$. Let the combination of step functions that define the region be denoted by $\Theta_A$, $\Theta_B$ and $\Theta_C$, for example $\Theta_A \equiv  \Theta(x - 1)\Theta(1 - y)\Theta(\theta_p  - 1)$. Then the total observable in each region can be written as,
	
	\begin{align}\label{eq:delta pt for a general clustering algorithm}\nn
		{\delta p_T}_{\text{ni}, A} &= p_T  z_1 (1 - \Xi_{\rm A}) \Theta_A,\\ 
		{\delta p_T}_{\text{ni}, B} &= -p_T  z_2 (1 - \Xi_{\rm B}) \Theta_B,\\\nn
		{\delta p_T}_{\text{ni}, C} &= -p_T  (z_1 + z_2) (1 - \Xi_{\rm C}) \Theta_C,
	\end{align}
	
	We now simply integrate the non-inclusive piece eq.(\ref{eq:non inclusive contribution to milan factor}) in these three regions using the corresponding observable, without the need to divide out calculation into many phase space regions as in previous calculations. Next, we will evaluate $\Xi_{\rm A}, \Xi_{\rm B}$ and $\Xi_{\rm C}$ for the C/A and $k_t$ algorithms.

	\subsection{A general jet-clustering algorithm}

	Generally, sequential clustering algorithms for events involving incoming hadrons are defined by a longitudinally invariant distance measure between the partons or jets in the event $d_{ij}$ and a beam distance $d_{iB}$, equipped with a jet radius $R$ \cite{Salam_2010},
	\begin{align}\label{eq:a general clustering algorithm}
		d_{ij} = \min(&{{\kappa^{2p}_{Ti}}, {\kappa^{2p}_{Tj}}})\frac{\Delta R^2_{ij}}{R^2},\\
		\nn
		d_{iB} = \ &{{\kappa^{2p}_{Ti}}},\\
		\nn
		\Delta R^2_{ij} = (y_i - y_j&)^2 + (\phi_i - \phi_j)^2.
	\end{align}

	$\kappa_{Ti}$ is the transverse momentum of parton $i$ with respect to the beam direction. For partons produced perpendicular to the beam this will be equivalent to the longitudinal momentum with respect to the jet $p_T z_i$. $y_i$ and $\phi_i$ are the rapidity and azimuth of the $i^{\text{th}}$ particle. In the collinear limit $\Delta R^2_{ij}$ is simply the small angle between the two particles $\theta_{ij}^2$. Therefore in the collinear limit we have
	\begin{align}\label{eq:a general clustering algorithm in the collinear limit}
		d_{ij} = p_T^{2p}\min(&z_i^{2p}, z_j^{2p})\frac{\theta_{ij}^2}{R^2},\\
		\nn
		d_{iB} = \ &p_T^{2p} z_i^{2p}.
	\end{align}
	
	\section{Results}
	In this section we will evaluate the algorithm clustering conditions $\Xi_{\rm A}, \Xi_{\rm B}$ and $\Xi_{\rm C}$ for each algorithm, and then substitute them into the general form for ${\delta p_T}_{\text{ni}}$, eq. \eqref{eq:delta pt for a general clustering algorithm}.
	\subsection{The observable in the non-inclusive case for the C/A algorithm}
	Unlike the $k_t$, anti-$k_t$ or unmodified Cambridge algorithm, there is no $k_t$ dependence in the distance measure in the C/A algorithm \cite{Dokshitzer:1997in,Wobisch:1998wt}. We set $p = 0$ in eq.\eqref{eq:a general clustering algorithm in the collinear limit}.  In the collinear limit $d_{ij}$ is simply the small angle between the two particles (or particle and jet) rescaled by the jet radius.
	\begin{align}\label{eq:cambridge aachen}
		d_{ij} = \frac{\theta^2_{ij}}{R^2},\\
		\nn
		d_{iB} = 1.
	\end{align}
	
	Partons are sequentially clustered by first finding the smallest $d_{ij}$. If it is smaller than the parton to beam distance, $d_{iB}$, partons $i$ and $j$ are clustered. $d_{iB}$ prevents clustering of partons a distance greater than $R$ from each other. In what follows, we will use our rescaled angular variables $x$, $y$, and $v$, namely
	\begin{equation}
		d_{1J} = x, \ d_{2J} = y, \ d_{12} = v.
	\end{equation}
	
	\subsubsection*{Region $A$}
	
	We get a clustering correction in region $A$ if $k_1$ and $k_2$ cluster, and so we require both $v$ < 1 and for $v$ to be the smallest distance measure $v < \min(x,y)$. In region A, $k_1$ is inside the jet and $k_2$ is outside the jet, therefore,
	\begin{equation}
		x < 1, \ y >1.
	\end{equation}
	
	Requiring $v < x$ then automatically ensures $v < 1$. This leads to the clustering condition
	
	\begin{equation}
		\Xi^{\rm CA}_{\rm A} = \Theta(x - v),
	\end{equation}
	and therefore the change in jet transverse momentum is given by
	\begin{align}\label{eq:delta pt for C/A in Region A}
		{\delta p_T}^{\rm CA}_{\text{ni},A} &= p_T z_1 (1 - \Theta(x - v)) \Theta_A,\\
		\nn
		&= p_T (1-z)z_p \Theta(v - x)\Theta_A.\\
		\nn
	\end{align} 
	\subsubsection*{Region $B$}
	The condition for clustering in region $B$ is the same as region $A$, and so
	\begin{align}\label{eq:delta pt for C/A in Region B}
		{\delta p_T}^{\rm CA}_{\text{ni},B} &= - p_T (1-z)(1-z_p) \Theta(v - x)\Theta_B.
	\end{align}
	
	\subsubsection*{Region $C$}
	
	In region $C$ both $k_1$ and $k_2$ are an angular distance greater than $R$ from the hard parton initiating the jet, so the condition for either to independently cluster to the jet cannot be met. But $k = k_1 + k_2$ is inside the jet, and so if $k_1$ and $k_2$ are clustered by the algorithm into $k$ first, the resulting entity will be within a radius of $R$ of the initiating hard parton and therefore capable of clustering to the jet. The condition for clustering is then simply $d_{12} < d_{iB}$, where $i = 1,2$. In other words $v < 1$, which gives us
	\begin{equation}
		\Xi^{\rm CA}_{\rm C} = \Theta(1 - v),
	\end{equation}
	and so the change is jet transverse momentum is given by
	\begin{align}\label{eq:delta pt for C/A in Region C}
		{\delta p_T}^{\rm CA}_{\text{ni},C} &= - p_T(1-z)\Theta(v - 1)\Theta_C. 
	\end{align}
	\subsection{The observable in the non-inclusive case for the $k_t$ algorithm}
	Setting $p=1$ in eq.\eqref{eq:a general clustering algorithm in the collinear limit}, we obtain the $k_t$ distance \cite{Catani:1993hr,Ellis:1993tq} 
	\begin{align}\label{eq:kt clustering algorithm in the collinear limit}
		d_{ij} = p_T^2\min(&z_i^2, z_j^2)\frac{\theta_{ij}^2}{R^2},\\
		\nn
		d_{iB} = \ &p_T^2 z_i^2.
	\end{align}
	\subsubsection*{Region $A$}
	Once again we get a correction if $k_1$ and $k_2$ recombine into $k$. This occurs if $d_{12} < d_{1J}$. Using eq.\eqref{eq:kt clustering algorithm in the collinear limit} this results in the condition $z_1^2 x > \min(z_1^2, z_2^2) v$, where recall $z_1 = z_p (1-z)$ and $z_2 = (1-z_p)(1-z)$ \eqref{eq:phase space variables}. This can be split into two terms, depending on if $z_p < 1/2$ or $z_p > 1/2$,
	\begin{equation}
		\Xi^{k_t}_{\rm A} = \Theta(x - v)\Theta(z_p < \frac{1}{2}) + \Theta(x - \frac{\left(1-z_p\right)^2}{z_p^2}v)\Theta(z_p > \frac{1}{2}),
	\end{equation}
	
	The change in jet transverse momentum in this region for the $k_t$ algorithm is then given by
	\begin{equation}\label{eq:region A kt}
		{\delta p_T}^{k_t}_{\text{ni},A} = p_T (1-z) z_p  \left\{\Theta(v - x)\Theta(z_p < \frac{1}{2}) + \Theta(\frac{\left(1-z_p\right)^2}{z_p^2}v - x)\Theta(z_p > \frac{1}{2})\right\}\Theta_A.
	\end{equation}
	\subsubsection*{Region $B$}
	Once again the condition for clustering in region $B$ is the same as region $A$, and so
	\begin{equation}\label{eq:region B kt}
		{\delta p_T}^{k_t}_{\text{ni},B} = - p_T (1-z)(1-z_p)  \left\{\Theta(v - x)\Theta(z_p < \frac{1}{2}) + \Theta(\frac{\left(1-z_p\right)^2}{z_p^2}v - x)\Theta(z_p > \frac{1}{2})\right\}\Theta_B.
	\end{equation}
	\subsubsection*{Region $C$}
	In the case of the $k_t$ algorithm, Region C has the same clustering condition as the C/A algorithm and hence the same value of the observable and phase space restrictions,
	\begin{align}\label{eq:region C kt}
		{\delta p_T}^{k_t}_{\text{ni},C} &= - p_T(1-z)\Theta(v - 1)\Theta_C.
	\end{align}
	\subsection{Massive gluon correction}\label{section:Massive gluon correction}
	As mentioned in Section \ref{section: the inclusive correction}, the Milan factor can be split into three terms. The naive case and the inclusive and non-inclusive corrections. In defining the latter two of pieces, we added and subtracted an inclusive term in the observable. This inclusive piece is dependent on the angle of the parent with respect to the jet, and therefore whether or not one defines the parent to be massive or massless. In  eq.\eqref{eq: relating parent angle to k_t}  we have assumed parent to be massless, for which we obtain
	\begin{align}
		\theta_{kJ} = \frac{\sqrt{k_T^2 + m^2}}{\alpha p_T} = \sqrt{\theta_{1J}^2 z_p + \theta_{2J}^2 (1-z_p)},
	\end{align}
	where here $m$ is the invariant mass of the two parton system. This is the definition used traditionally in Milan factor calculations. A massive parent would instead use the definition
	\begin{align}
		\theta_{kJ} = \frac{k_T}{\alpha p_T} = \sqrt{\theta_{1J}^2 z_p + \theta_{2J}^2 (1-z_p) - \theta_{12}^2 z_p (1-z_p)}.
	\end{align}

	We have used the second definition throughout our calculation of the non-inclusive correction. But as mentioned, the original calculation \cite{Dokshitzer_1998} used the first definition in both the non-inclusive and inclusive correction. We would like to simply quote the result for the inclusive case from the original article and focus on the non-inclusive correction as that is where the clustering dependence lies. In order to do this, we must correct our answer. This amounts to changing the definition of $\theta_p$ when calculating the cone case
	\begin{equation}\label{eq:definition of parent angle}
		\theta^2_p = \frac{\theta^2_{kJ}}{R^2} = x z_p + y (1-z_p) - v z_p (1-z_p) \rightarrow \tilde{\theta}^2_p = x z_p + y(1-z_p).
	\end{equation}
	
	In regions $A,B$ and $C$ the corrections would be
	\begin{equation}\label{eq:cone correction region A}
		{\delta p_T}^{\rm correction}_{\text{cone},A} = p_T  z_1 \Theta(x - 1)\Theta(1 - y)\left\{\Theta(\tilde{\theta}_p -1) - \Theta(\theta_p  - 1)\right\},
	\end{equation}
	\begin{equation}\label{eq:cone correction region B}
		{\delta p_T}^{\rm correction}_{\text{cone},B} = - p_T  z_2 \Theta(x - 1)\Theta(1 - y)\left\{\Theta(\tilde{\theta}_p -1) - \Theta(\theta_p  - 1)\right\},
	\end{equation}
	\begin{equation}\label{eq:cone correction region C}
		{\delta p_T}^{\rm correction}_{\text{cone},C} = - p_T  (z_1 + z_2) \Theta(x - 1)\Theta(1 - y)\left\{\Theta(\tilde{\theta}_p -1) - \Theta(\theta_p  - 1)\right\}.
	\end{equation}	
	\subsection{Numerical integration of $\delta p_T$}
	In this section we present results for the non-universal Milan factor in the case of the C/A algorithm, and also check to see if we reproduce the known $k_t$ result.
	
	\subsubsection{C/A result}
	Recall we are calculating eq.\eqref{eq:non inclusive contribution to milan factor}. It is standard to divide the total matrix element squared into three pieces, namely equations \eqref{eq:qqbar collinear splitting in soft limit}, \eqref{eq:leading infrared contribution in collinear limit} and \eqref{eq:finite remainder of non-abelian contribution in the soft limit}. We will go into detail for the first case, in which the gluon decays into a quark-antiquark pair. We have 
	\begin{align}\label{eq:non-inclusive case region nf piece}
		{r^{(n)}}^{\rm CA}_{\text{$n_f$}} = \frac{\sqrt{\rho}}{2 \beta_0 p_T}  \ \int  d \Gamma_2 \ 	\langle \hat{P}_{\bar{q}'_1 q'_2 q_3} \rangle \ \delta {p_T}^{\rm CA}_{\text{ni}}  \ \delta(\rho - z_p (1-z_p)(1-z)^2 v),
	\end{align}
	\begin{align}
		\ \delta {p_T}^{\rm CA}_{\text{ni}} &= {\delta p_T}^{\rm CA}_{\text{ni},A} + {\delta p_T}^{\rm CA}_{\text{ni},B} + {\delta p_T}^{\rm CA}_{\text{ni},C},
	\end{align}
	\begin{align}
		d \Gamma_2 &= dxdydv \Delta^{-\frac{1}{2}} J_1,
	\end{align}
	where $\Delta$, $J_1$ are given in eq.\eqref{eq:triple collinear phase space factors} and $x,y,z$ are the rescaled angles defined in eq.\eqref{eq:defining x y and v}.
	
	The subscript on ${r^{(n)}}^{\rm CA}_{\text{$n_f$}}$ denotes the fact this term is proportional to the number of flavours $n_f$. We will refer to this as the $n_f$ piece. The other terms are related to the gluon splitting into two gluons and are proportional to $C_A$.
	\subsubsection*{Region $A$}
	We substitute into the above equations \eqref{eq:triple collinear phase space factors}, \eqref{eq:qqbar collinear splitting in soft limit}, and only retain the contribution to the observable from region A, namely eq.\eqref{eq:delta pt for C/A in Region A}. After fixing $z$ using the delta function, we have
	\begin{align}\label{eq:region A for the nf term, CA algorithm}
		{r^{(n)}}^{\rm CA}_{\text{$n_f$,$A$}} = - \frac{1}{\beta_0}\frac{n_f}{\pi}  \ \int  dx dy   dv \Delta^{-\frac{1}{2}} \  \frac{ (x-y)^2 z_p(1-z_p) - v\left(x z_p +y(1-z_p)\right)}{v^2 (x z_p + y (1-z_p))^2 \sqrt{z_p (1-z_p)}} \  z_p  \Theta(v - x)\Theta(\Delta)\Theta_A.
	\end{align}
	
	The factors of $p_T$ and $\rho$ cancel out leaving a constant. Recall $\Theta(\Delta)$ comes about due to the positivity of the Gram determinant and this imposes limits on $v$
	\begin{align}
		(\sqrt{x} - \sqrt{y})^2 < v < (\sqrt{x} + \sqrt{y})^2,
	\end{align} 
	and $\Theta_A = \Theta(x-1)\Theta(1-y)\Theta(\theta_p-1)$ restricts us to region $A$. $\theta_p$ is given by the first relation in eq.\eqref{eq:definition of parent angle}. Upon integration we obtain
	\begin{align}\label{eq:region A nf piece}
		{r^{(n)}}^{\rm CA}_{\text{$n_f$,$A$}}  = - \frac{1}{\beta_0} 0.0383 n_f.
	\end{align}
	\subsubsection*{Regions $B$ and $C$}
	We repeat the above calculation, but instead using \eqref{eq:delta pt for C/A in Region B} for region $B$ and \eqref{eq:delta pt for C/A in Region C} for region $C$. We obtain the results
	\begin{align}\label{eq:region B nf piece}
		{r^{(n)}}^{\rm CA}_{\text{$n_f$,$B$}}  = \frac{1}{\beta_0} 0.0963 n_f,
	\end{align}
	\begin{align}\label{eq:region C nf piece}
		{r^{(n)}}^{\rm CA}_{\text{$n_f$,$C$}}  = \frac{1}{\beta_0} 0.0881 n_f.
	\end{align}
	\subsubsection*{The massive gluon correction}
	The massive gluon correction is then simply an extension of the above calculation. We correct for our massless treatment of the inclusive piece by instead substituting in for our observable eq.\eqref{eq:cone correction region A} in region $A$, eq.\eqref{eq:cone correction region B} in region $B$ and eq.\eqref{eq:cone correction region C} in region $C$. Doing so, we obtain
	\begin{align}\label{eq:region A correction nf piece}
		{r^{(n)}}^{\rm CA \ correction}_{\text{$n_f$,$A$}} = -\frac{1}{\beta_0} 0.0812 n_f,
	\end{align}
	\begin{align}\label{eq:region B correction nf piece}
		{r^{(n)}}^{\rm CA \ correction}_{\text{$n_f$,$B$}} = -\frac{1}{\beta_0} 0.0591 n_f,
	\end{align}
	\begin{align}\label{eq:region C correction nf piece}
		{r^{(n)}}^{\rm CA \ correction}_{\text{$n_f$,$C$}} = -\frac{1}{\beta_0} 0.0967 n_f.
	\end{align}
	\subsubsection*{Non-inclusive correction for the $n_f$ piece}
	Recall that contributions from regions $A$ and $B$ must be accompanied by a factor of 2 in order to account for the possibility of interchanging the decay products $k_1$ and $k_2$. Summing up the contributions eq.(\ref{eq:region A nf piece}-\ref{eq:region C correction nf piece})
	\begin{align}
		{r^{(n)}}^{\rm CA}_{\text{$n_f$}} = -\frac{2}{\beta_0} 0.087 n_f.
	\end{align}

	\subsubsection*{Full non-inclusive C/A correction}
	Repeating the above exercise using eq.\eqref{eq:leading infrared contribution in collinear limit} and eq.\eqref{eq:finite remainder of non-abelian contribution in the soft limit} as our matrix element squared we obtain the full result for the non-inclusive C/A correction 
	\begin{align}
		{r^{(n)}}^{\rm CA} = \frac{2}{\beta_0}\left(-1.835 C_A + 0.519 C_A - 0.087 n_f\right).
	\end{align}

	We substitute this term, along with the inclusive correction eq.\eqref{eq:inclusive correction} into eq.\eqref{eq:milan factor definition} to produce an overall factor one can use to correct for the naive inclusive treatment eq.\eqref{eq:naive answer},
	\begin{equation}
		\langle\delta p_T\rangle^{\rm CA} = -\frac{2 A_1}{R} \left(1 + {r^{(i)}} + {r^{(n)}}^{\rm CA}\right) = -\frac{2 A_1}{R} \mathcal{M_{\rm CA}},
	\end{equation}
	where $A_1$ is given by eq.\eqref{eq:npt moment of coupling in new variables}. Using $C_A = 3, n_f = 3$ and quoting the known algorithm independent inclusive piece from \cite{Dokshitzer_1998} $r^{(i)} = 3.299 \frac{C_A}{\beta_0}$ we find the non-universal Milan factor
	\begin{equation}
		\mathcal{M}^{\rm CA}= 1.16.
	\end{equation}
	\subsubsection{$k_t$ result}
	\subsubsection*{Full non-inclusive $k_t$ correction}
	Repeating the exercise above but instead using observable after applying the $k_t$ algorithm eq.\eqref{eq:region A kt}, eq.\eqref{eq:region B kt} and eq.\eqref{eq:region C kt}, we obtain
	\begin{align}
		{r^{(n)}}^{k_t} = \frac{2}{\beta_0}(-2.140 C_A + 0.608 C_A  -0.096 n_f)
	\end{align}
	which within numerical integration error is consistent with \cite{Dasgupta:2009tm}. As a result we recover the known non-universal Milan factor for the $k_t$ algorithm
	\begin{equation}
		\mathcal{M}^{k_t} = 1.01.
	\end{equation}
	
	\section{Conclusions and future prospects}	
	In performing an $\mathcal{O}\left(\alpha_s\right)$ analysis of non-perturbative effects, one can exploit the correspondence between renormalons and perturbative calculations performed with a finite gluon mass in order to quantify hadronization corrections. 
	However, a two-loop analysis is at the very least necessary to quantify the scale at which non-perturbative physics breaks down. In addition, there are further issues that render a massive single gluon approach insufficient for determining the magnitude of power corrections. There is an inherent ambiguity brought about by the inclusion of small gluon mass effects, namely how to include the gluon mass consistently throughout the calculation \cite{Dasgupta_1998, Dokshitzer_1996}. As well as this, a one-loop massive gluon approach entails inclusively integrating over the gluon decay products. This does not account for a complete description of the 4-parton final state kinematics that some observables rely on in order to predict accurately the magnitude of their power correction \cite{Nason:1995np, Dokshitzer_1998, Dokshitzer_1998_2}. One must extend the analysis to two-loop accuracy in order to resolve these problems.
	
	While an $\mathcal{O}\left(\alpha_s^2\right)$ analysis has been carried out for the $1/Q$ power corrections to jet shape variables, the universal Milan factor that is used to correct for a naive one-loop analysis of jet shapes can only be applied to observables that depend linearly on the final state momenta.
	
	Observables calculated with a jet-clustering step must be treated more carefully. While the rigid cone algorithm only depends linearly on the final state momenta, practical algorithms used in phenomenology, such as the $k_t$, C/A and anti-$k_t$ algorithms, involve sequential clustering and hence depend non-linearly on final state momenta.
	
	In this paper we have used a dispersive method to carry out an $\mathcal{O}\left(\alpha_s^2\right)$ analysis of the leading hadronic correction to the change in jet transverse momentum after applying a jet clustering algorithm in the small $R$ limit. We restricted ourselves to the soft limit which is responsible for hadronization corrections, and the triple-collinear limit which is sufficient to capture the leading $1/R$ hadronization effect. This approach allowed us to define the inclusive and non-inclusive corrections \cite{Dokshitzer_1998} necessary to accurately predict the magnitude of the power correction. We focused on the non-inclusive correction as it is the only contribution dependent on the clustering algorithm. For the inclusive correction, we quoted the known result \cite{Dokshitzer_1998}.
	
	We have also utilised an alternate phase-space parametrisation in which one can easily divide the phase into three regions intuitive for calculations that include a jet clustering step. This is because the regions are ultimately defined by whether a particular final state particle is inside or outside the jet, which then defines the observable in that region.
	
	In this paper we have found a new result for the Cambridge Aachen algorithm. We have derived an $\mathcal{O}(1)$  correction factor, analogous to the Milan factor for jet shapes, required to correct for the inadequacies of a one-loop analysis. This is given by $\mathcal{M}^{\rm CA}= 1.16$. 	As a cross-check, we also re-calculated the $k_t$ case and independently reproduced the results of \cite{Dasgupta:2009tm} with our new phase space parametrisation, confirming $\mathcal{M}^{k_t}= 1.01.$
	
	Our result for the C/A algorithm is important since it is the primary choice of clustering algorithm for the application of jet substructure techniques e.g. soft drop, in which the groomed jet mass is sensitive to changes in sub-jet $p_t$ \cite{Dasgupta:2013ihk}. Such techniques are vital in LHC phenomenology, and have also been used for the extraction of $\alpha_s$. State of the art $\alpha_s$ extraction has set high standards on the level of accuracy necessary for a competitive extraction  \cite{Nason:2023asn}. To achieve the required accuracy, precise predictions of the magnitude of hadronization corrections to such quantities are of importance.
	
	It is to be hoped that our work in the present article will be useful for future more accurate studies of corrections for observables that involve a clustering step.

	\subsection*{Acknowledgements}
	I would like to thank Mrinal Dasgupta and Aditya Pathak for discussions which initiated this study, and their continued guidance. I would also like to thank the CERN theory department for their hospitality during the course of this research. Finally I would like to thank the UK Science and Technology Facilities Council for funding my PhD studentship.

	\begin{appendices}
		
		\section{Cambridge Aachen under previous parametrisation}\label{Cambridge Aachen old method}
		Here we state the clustering conditions and non-inclusive trigger functions for the CA case, and recalculate the result using the method outlined in \cite{Dasgupta:2009tm}.
		The clustering function is given by
		\begin{align}\label{eq:CACluster}
			\Xi^{\text{CA}}_{\rm out} (k_1) &= \Theta_{\rm out}(k_1)
			\nn \\
			&-\Theta_{\rm out}(k_1)\Theta_{\rm out}(k_2)\Theta_{12}(k_1,k_2)\Theta(R^2 - \theta^2_{kJ})\big]
			\nn \\
			&+ \Theta_{\rm in}(k_1)\Theta_{\rm out}(k_2)\big[\Theta_{12}(k_1,k_2)
			\Theta(\theta^2_{Jk} - R^2) \Theta(\theta^2_{1J} - \theta^2_{12})\big]
			\nn \\
			&-
			\Theta_{\rm out}(k_1)\Theta_{\rm in}(k_2)\Theta_{12}(k_1,k_2)\Theta(\theta^2_{2J} - \theta^2_{12})\Theta(R^2 - \theta^2_{Jk}) .
		\end{align}
		
		After integrating over the gluon mass $\alpha$ (what we labelled $1-z$), we obtain
		\begin{align}\label{eq:Integrated noninclusive trigger function in CA scheme}
			\Omega_{\rm ni }=& \frac{1}{R}\big[k_{T_1} + k_{T_2} - \sqrt{k_T^2 + m^2} + zf^{\text{CA}}(q_1, q_2) + (1-z)f^{\text{CA}}(q_2,q_1)\big],
		\end{align}
		where the function $f^{\rm CA}$ encapsulates the algorithm dependent contribution. It is given by 
		\begin{align}\label{eq:fCA}
			f^{\text{CA}}(q_1,q_2) \equiv \ - &\Psi (\min(q_1,q_2) - \max(q,k_T))
			\nn \\
			+ &\Psi (\min(q_2, k_T) - \max(q_1,q))\Theta(q_1-q)
			\nn \\
			- &\Psi (q_1 - \max(q_2,q,k_T))\Theta(q_2-q).
		\end{align}
		
		In the above we have adopted the notation of \cite{Dasgupta:2009tm}, $\Psi(x) \equiv x \Theta(x)$. One then defines $u_i \equiv \frac{q_i}{q}$, $\tilde{\Omega}_{\rm ni }(u_1,u_2) \equiv \Omega_{\rm ni }(q_1,q_2) \frac{R}{q}$ and $X \equiv \sqrt{z u_1^2 + (1-z) u_2^2 -z(1-z)}$. One assumes $u_1 > u_2$ and doubles the answer at the end. Separating the trigger function into the various regions of phase space, we obtain purely the clustering correction which we label as $\tilde{\Omega}_{\rm diff }$
		\begin{align}\label{eq:phase space splitting CA}
			1 < X < u_2 & \implies \tilde{\Omega}_{\rm diff } = \big[X - z u_1 -(1-z)u_2\big]\Theta(u_2 - X),
			\nn \\
			1  < u_2 < X  & \implies \tilde{\Omega}_{\rm diff } = \big[X - z u_1 -(1-z)u_2\big]\Theta(X - u_2),
			\nn \\
			u_2  < X < 1  & \implies \tilde{\Omega}_{\rm diff } = 0,
			\nn \\
			X  < u_2 < 1  & \implies \tilde{\Omega}_{\rm diff } = 0,
			\nn \\
			u_2 < 1 < X  & \implies \tilde{\Omega}_{\rm diff } = 0,
			\nn \\
			X < 1 < u_2  & \implies \tilde{\Omega}_{\rm diff } = \big[1- z u_1 - (1-z)u_2\big].
		\end{align}

		Comparing to \cite{Dasgupta:2009tm}, The primary difference between the clustering functions of the $k_t$ and C/A algorithms is that $\Omega(u_2 - \min(1,\frac{z}{1-z})) \rightarrow \Omega(u_2 - 1))$. As a result, in addition to $u_1 < 1$ collapsing any potential clustering from the algorithm, so does $u_2 < 1$. Regions $A$, $\tilde{A}$ and $D$ remain the same. Regions $B$ and $C$ now have no corrections from clustering, similar to the phase space region $D$.
		
		One now calculates the non inclusive correction using the above trigger function
		\begin{align}\label{eq:non inclusive correction old}
			{r^{(n)}} = \frac{1}{\pi \beta_0} \int_0^1  \frac{dz}{\sqrt{z(1-z)}}\int_0^{\infty} du_1 u_1 \int_0^{\infty} du_2 u_2 \frac{\mathcal{M}^2}{z u_1 ^2 + (1-z) u_2^2} \frac{\tilde{\Omega}_{\rm ni}}{\sqrt{J(u_1,u_2)}}  \Theta(J(u_1,u_2)),
		\end{align}
		$\mathcal{M}^2$ is defined as the non-abelian contribution to the matrix element squared rescaled such that it is written purely in terms of the $u_i$ variables,
		\begin{align}\label{eq:define S H_g and H_q}
			\mathcal{M}^2(u_1,u_2) &= 2 C_A(2\mathcal{S} + \mathcal{H}_g) + 2n_f\mathcal{H}_q
			\nn \\
			&=  [2 C_A(2 S + H_g) + 2n_f H_q]m^2(zq_1^2 + (1-z)q_2^2)
			\nn \\
			&= M^2(k_1,k_2)m^2(zq_1^2 + (1-z)q_2^2),
		\end{align}
		$\mathcal{S}$ is the leading infrared contribution of the non abelian piece while the remainder $\mathcal{H}_q, \mathcal{H}_g$ are related to the parent gluon decaying into quarks and gluons respectively of the same order. $J$ is the Jacobian factor resulting from changing the transverse and azimuthal phase space variables to the $u_i$ variables
		\begin{align}
			J \equiv ((u_1 + u_2)^2 - 1)(1 - (u_1 - u_2)^2),
		\end{align}
		and the $\tilde{\Omega}_{\rm ni}$ is the non-inclusive trigger function. In the case of clustering corrections, as explained in this article, the naive and inclusive trigger functions will not change between different clustering algorithms. As a result, this is the only contribution we will consider.\\
		\begin{align}
			{r^{(n)}}^{\rm CA} = \frac{2}{\beta_0}(&-1.227 C_A + 0.365 C_A - 0.052 n_f
			\nn \\ &-0.610 C_A + 0.154 C_A - 0.035 n_f
			\nn \\ &-0.003 C_A + 0.001 C_A - 0.000 n_f),
		\end{align}
		combining this with the inclusive correction $r^{(i)}$  \cite{Dokshitzer_1998}, which by definition is unchanged by clustering, for $C_A = 3$, $n_f = 3$,
		\begin{align}
			\langle \delta p_t \rangle &= -2 \frac{A_1}{R}(1 + r^{(i)} + r^{(n)})
			\nn \\
			&= -2 \frac{A_1}{R}(1.160),
		\end{align}
		where $A_1$ is the first moment of the non-perturbative coupling
		\begin{align}
			A_1 = \frac{C_F}{2\pi}\int_{0}^{\infty} \frac{dm^2}{m^2}m \delta \alpha_{\rm eff}(m^2).
		\end{align}
		\section{A dictionary between the two methods}
		We instead write our two-body phase space as
		\begin{align}
			d \Gamma_2 = d \Gamma_2(1-z, \frac{\theta_{12}}{R}, \frac{\theta_{13}}{R}, \frac{\theta_{23}}{R}, z_p),
		\end{align}
		where $R$ is the jet radius. A dictionary between the two sets of integration variables is as follows
		\begin{align*}
			\alpha &= (1-z),
		\end{align*}
		\begin{align}
			u_1 \equiv \frac{\abs{\bar{q}_1}}{q} &= \frac{\theta_{13}}{\theta_{12}}, \ \ u_2 \equiv \frac{\abs{\bar{q}_2}}{q} = \frac{\theta_{23}}{\theta_{12}},
		\end{align}
		\begin{align*}
			u_+ &\equiv u_1 + u_2 ,	 \ \ u_- \equiv u_1 - u_2,
		\end{align*}
		\begin{align*}
			q &\equiv \abs{\bar{q}_1 - \bar{q}_2} = (1-z)p_T \theta_{12}.
		\end{align*}

		Now $z$ represents the energy fraction retained by the quark emitting the gluon, hence $1-z$ is the fraction of energy the gluon takes from the quark. $z_p$ and $1-z_p$ represent the energy fractions the decay products take from the parent gluon now that $z$ has been repurposed.
		
		As a reminder, $q_i$ were the rescaled transverse momenta with respect to the jet,  $ \frac{k_{T_i}}{z_i} $, hence the connection to the angles between said partons and the hard parton initiation the jet $\theta_{i3}$. Additionally, the dictionary above is only valid in the collinear limit.

	\end{appendices}
	\newpage
	\bibliography{../qcd}

\providecommand{\href}[2]{#2}\begingroup\raggedright\begin{thebibliography}{10}

\bibitem{Dasgupta:2003iq}
M.~Dasgupta and G.~P. Salam, \emph{{Event shapes in e+ e- annihilation and deep
  inelastic scattering}},
  \href{https://doi.org/10.1088/0954-3899/30/5/R01}{\emph{J. Phys. G}
  {\bfseries 30} (2004) R143}
  [\href{https://arxiv.org/abs/hep-ph/0312283}{{\ttfamily hep-ph/0312283}}].

\bibitem{Gehrmann-DeRidder:2007nzq}
A.~Gehrmann-De~Ridder, T.~Gehrmann, E.~W.~N. Glover and G.~Heinrich,
  \emph{{Second-order QCD corrections to the thrust distribution}},
  \href{https://doi.org/10.1103/PhysRevLett.99.132002}{\emph{Phys. Rev. Lett.}
  {\bfseries 99} (2007) 132002}
  [\href{https://arxiv.org/abs/0707.1285}{{\ttfamily 0707.1285}}].

\bibitem{Dissertori:2008cn}
G.~Dissertori, A.~Gehrmann-De~Ridder, T.~Gehrmann, E.~W.~N. Glover, G.~Heinrich
  and H.~Stenzel, \emph{{e+ e- ---\ensuremath{>} 3 jets and event shapes at
  NNLO}}, \href{https://doi.org/10.1016/j.nuclphysbps.2008.09.072}{\emph{Nucl.
  Phys. B Proc. Suppl.} {\bfseries 183} (2008) 2}
  [\href{https://arxiv.org/abs/0806.4601}{{\ttfamily 0806.4601}}].

\bibitem{Gehrmann-DeRidder:2009fgd}
A.~Gehrmann-De~Ridder, T.~Gehrmann, E.~W.~N. Glover and G.~Heinrich,
  \emph{{NNLO moments of event shapes in e+e- annihilation}},
  \href{https://doi.org/10.1088/1126-6708/2009/05/106}{\emph{JHEP} {\bfseries
  05} (2009) 106} [\href{https://arxiv.org/abs/0903.4658}{{\ttfamily
  0903.4658}}].

\bibitem{Weinzierl:2008iv}
S.~Weinzierl, \emph{{NNLO corrections to 3-jet observables in electron-positron
  annihilation}},
  \href{https://doi.org/10.1103/PhysRevLett.101.162001}{\emph{Phys. Rev. Lett.}
  {\bfseries 101} (2008) 162001}
  [\href{https://arxiv.org/abs/0807.3241}{{\ttfamily 0807.3241}}].

\bibitem{Weinzierl:2009yz}
S.~Weinzierl, \emph{{Moments of event shapes in electron-positron annihilation
  at NNLO}}, \href{https://doi.org/10.1103/PhysRevD.80.094018}{\emph{Phys. Rev.
  D} {\bfseries 80} (2009) 094018}
  [\href{https://arxiv.org/abs/0909.5056}{{\ttfamily 0909.5056}}].

\bibitem{Gehrmann:2019hwf}
T.~Gehrmann, A.~Huss, J.~Mo and J.~Niehues, \emph{{Second-order QCD corrections
  to event shape distributions in deep inelastic scattering}},
  \href{https://doi.org/10.1140/epjc/s10052-019-7528-3}{\emph{Eur. Phys. J. C}
  {\bfseries 79} (2019) 1022}
  [\href{https://arxiv.org/abs/1909.02760}{{\ttfamily 1909.02760}}].

\bibitem{Alvarez:2023fhi}
M.~Alvarez, J.~Cantero, M.~Czakon, J.~Llorente, A.~Mitov and R.~Poncelet,
  \emph{{NNLO QCD corrections to event shapes at the LHC}},
  \href{https://doi.org/10.1007/JHEP03(2023)129}{\emph{JHEP} {\bfseries 03}
  (2023) 129} [\href{https://arxiv.org/abs/2301.01086}{{\ttfamily
  2301.01086}}].

\bibitem{Stewart_2011}
I.~W. Stewart, F.~J. Tackmann and W.~J. Waalewijn, \emph{Beam thrust cross
  section for drell-yan production at next-to-next-to-leading-logarithmic
  order}, \href{https://doi.org/10.1103/physrevlett.106.032001}{\emph{Physical
  Review Letters} {\bfseries 106} (2011) }.

\bibitem{Jouttenus_2013}
T.~T. Jouttenus, I.~W. Stewart, F.~J. Tackmann and W.~J. Waalewijn, \emph{Jet
  mass spectra in higgs boson plus one jet at next-to-next-to-leading
  logarithmic order},
  \href{https://doi.org/10.1103/physrevd.88.054031}{\emph{Physical Review D}
  {\bfseries 88} (2013) }.

\bibitem{Kang:2013nha}
D.~Kang, C.~Lee and I.~W. Stewart, \emph{{Using 1-Jettiness to Measure 2 Jets
  in DIS 3 Ways}},
  \href{https://doi.org/10.1103/PhysRevD.88.054004}{\emph{Phys. Rev. D}
  {\bfseries 88} (2013) 054004}
  [\href{https://arxiv.org/abs/1303.6952}{{\ttfamily 1303.6952}}].

\bibitem{Banfi:2014sua}
A.~Banfi, H.~McAslan, P.~F. Monni and G.~Zanderighi, \emph{{A general method
  for the resummation of event-shape distributions in $e^{+} e^{−}$
  annihilation}}, \href{https://doi.org/10.1007/JHEP05(2015)102}{\emph{JHEP}
  {\bfseries 05} (2015) 102} [\href{https://arxiv.org/abs/1412.2126}{{\ttfamily
  1412.2126}}].

\bibitem{Becher:2015gsa}
T.~Becher and X.~Garcia~i Tormo, \emph{{Factorization and resummation for
  transverse thrust}},
  \href{https://doi.org/10.1007/JHEP06(2015)071}{\emph{JHEP} {\bfseries 06}
  (2015) 071} [\href{https://arxiv.org/abs/1502.04136}{{\ttfamily
  1502.04136}}].

\bibitem{Banfi:2016zlc}
A.~Banfi, H.~McAslan, P.~F. Monni and G.~Zanderighi, \emph{{The two-jet rate in
  $e^+e^-$ at next-to-next-to-leading-logarithmic order}},
  \href{https://doi.org/10.1103/PhysRevLett.117.172001}{\emph{Phys. Rev. Lett.}
  {\bfseries 117} (2016) 172001}
  [\href{https://arxiv.org/abs/1607.03111}{{\ttfamily 1607.03111}}].

\bibitem{Becher_2016}
T.~Becher, X.~G. i~Tormo and J.~Piclum, \emph{Next-to-next-to-leading
  logarithmic resummation for transverse thrust},
  \href{https://doi.org/10.1103/physrevd.93.054038}{\emph{Physical Review D}
  {\bfseries 93} (2016) }.

\bibitem{Tulipant:2017ybb}
Z.~Tulip\'ant, A.~Kardos and G.~Somogyi, \emph{{Energy\textendash{}energy
  correlation in electron\textendash{}positron annihilation at NNLL + NNLO
  accuracy}}, \href{https://doi.org/10.1140/epjc/s10052-017-5320-9}{\emph{Eur.
  Phys. J. C} {\bfseries 77} (2017) 749}
  [\href{https://arxiv.org/abs/1708.04093}{{\ttfamily 1708.04093}}].

\bibitem{Banfi:2018mcq}
A.~Banfi, B.~K. El-Menoufi and P.~F. Monni, \emph{{The Sudakov radiator for jet
  observables and the soft physical coupling}},
  \href{https://doi.org/10.1007/JHEP01(2019)083}{\emph{JHEP} {\bfseries 01}
  (2019) 083} [\href{https://arxiv.org/abs/1807.11487}{{\ttfamily
  1807.11487}}].

\bibitem{Gao:2019ojf}
A.~Gao, H.~T. Li, I.~Moult and H.~X. Zhu, \emph{{Precision QCD Event Shapes at
  Hadron Colliders: The Transverse Energy-Energy Correlator in the Back-to-Back
  Limit}}, \href{https://doi.org/10.1103/PhysRevLett.123.062001}{\emph{Phys.
  Rev. Lett.} {\bfseries 123} (2019) 062001}
  [\href{https://arxiv.org/abs/1901.04497}{{\ttfamily 1901.04497}}].

\bibitem{Kardos:2020gty}
A.~Kardos, A.~J. Larkoski and Z.~Tr\'ocs\'anyi, \emph{{Groomed jet mass at high
  precision}},
  \href{https://doi.org/10.1016/j.physletb.2020.135704}{\emph{Phys. Lett. B}
  {\bfseries 809} (2020) 135704}
  [\href{https://arxiv.org/abs/2002.00942}{{\ttfamily 2002.00942}}].

\bibitem{Dasgupta:2022fim}
M.~Dasgupta, B.~K. El-Menoufi and J.~Helliwell, \emph{{QCD resummation for
  groomed jet observables at NNLL+NLO}},
  \href{https://doi.org/10.1007/JHEP01(2023)045}{\emph{JHEP} {\bfseries 01}
  (2023) 045} [\href{https://arxiv.org/abs/2211.03820}{{\ttfamily
  2211.03820}}].

\bibitem{vanBeekveld:2023lsa}
M.~van Beekveld, M.~Dasgupta, B.~K. El-Menoufi, J.~Helliwell and P.~F. Monni,
  \emph{{Collinear fragmentation at NNLL: generating functionals, groomed
  correlators and angularities}},
  \href{https://arxiv.org/abs/2307.15734}{{\ttfamily 2307.15734}}.

\bibitem{Chen:2023zlx}
W.~Chen, J.~Gao, Y.~Li, Z.~Xu, X.~Zhang and H.~X. Zhu, \emph{{NNLL Resummation
  for Projected Three-Point Energy Correlator}},
  \href{https://arxiv.org/abs/2307.07510}{{\ttfamily 2307.07510}}.

\bibitem{Bhattacharya:2023qet}
A.~Bhattacharya, J.~K.~L. Michel, M.~D. Schwartz, I.~W. Stewart and X.~Zhang,
  \emph{{NNLL Resummation of Sudakov Shoulder Logarithms in the Heavy Jet Mass
  Distribution}},  \href{https://arxiv.org/abs/2306.08033}{{\ttfamily
  2306.08033}}.

\bibitem{Becher:2008cf}
T.~Becher and M.~D. Schwartz, \emph{{A precise determination of $\alpha_s$ from
  LEP thrust data using effective field theory}},
  \href{https://doi.org/10.1088/1126-6708/2008/07/034}{\emph{JHEP} {\bfseries
  07} (2008) 034} [\href{https://arxiv.org/abs/0803.0342}{{\ttfamily
  0803.0342}}].

\bibitem{Chien:2010kc}
Y.-T. Chien and M.~D. Schwartz, \emph{{Resummation of heavy jet mass and
  comparison to LEP data}},
  \href{https://doi.org/10.1007/JHEP08(2010)058}{\emph{JHEP} {\bfseries 08}
  (2010) 058} [\href{https://arxiv.org/abs/1005.1644}{{\ttfamily 1005.1644}}].

\bibitem{Abbate:2010xh}
R.~Abbate, M.~Fickinger, A.~H. Hoang, V.~Mateu and I.~W. Stewart, \emph{{Thrust
  at $N^{3}LL$ with Power Corrections and a Precision Global Fit for
  $\alpha_{s}(mZ)$}},
  \href{https://doi.org/10.1103/PhysRevD.83.074021}{\emph{Phys. Rev. D}
  {\bfseries 83} (2011) 074021}
  [\href{https://arxiv.org/abs/1006.3080}{{\ttfamily 1006.3080}}].

\bibitem{Hoang:2014wka}
A.~H. Hoang, D.~W. Kolodrubetz, V.~Mateu and I.~W. Stewart,
  \emph{{$C$-parameter distribution at N$^3$LL' including power corrections}},
  \href{https://doi.org/10.1103/PhysRevD.91.094017}{\emph{Phys. Rev. D}
  {\bfseries 91} (2015) 094017}
  [\href{https://arxiv.org/abs/1411.6633}{{\ttfamily 1411.6633}}].

\bibitem{Sjostrand:2006za}
T.~Sjostrand, S.~Mrenna and P.~Z. Skands, \emph{{PYTHIA 6.4 Physics and
  Manual}}, \href{https://doi.org/10.1088/1126-6708/2006/05/026}{\emph{JHEP}
  {\bfseries 05} (2006) 026}
  [\href{https://arxiv.org/abs/hep-ph/0603175}{{\ttfamily hep-ph/0603175}}].

\bibitem{Winter:2003tt}
J.-C. Winter, F.~Krauss and G.~Soff, \emph{{A Modified cluster hadronization
  model}}, \href{https://doi.org/10.1140/epjc/s2004-01960-8}{\emph{Eur. Phys.
  J. C} {\bfseries 36} (2004) 381}
  [\href{https://arxiv.org/abs/hep-ph/0311085}{{\ttfamily hep-ph/0311085}}].

\bibitem{Sherpa:2019gpd}
{\scshape Sherpa} collaboration, E.~Bothmann et~al., \emph{{Event Generation
  with Sherpa 2.2}},
  \href{https://doi.org/10.21468/SciPostPhys.7.3.034}{\emph{SciPost Phys.}
  {\bfseries 7} (2019) 034} [\href{https://arxiv.org/abs/1905.09127}{{\ttfamily
  1905.09127}}].

\bibitem{Kupco:1998fx}
A.~Kupco, \emph{{Cluster hadronization in HERWIG 5.9}},  in \emph{{Workshop on
  Monte Carlo Generators for HERA Physics (Plenary Starting Meeting)}},
  pp.~292--300, 4, 1998, \href{https://arxiv.org/abs/hep-ph/9906412}{{\ttfamily
  hep-ph/9906412}}.

\bibitem{Bahr:2008pv}
M.~Bahr et~al., \emph{{Herwig++ Physics and Manual}},
  \href{https://doi.org/10.1140/epjc/s10052-008-0798-9}{\emph{Eur. Phys. J. C}
  {\bfseries 58} (2008) 639} [\href{https://arxiv.org/abs/0803.0883}{{\ttfamily
  0803.0883}}].

\bibitem{Webber}
B.~Webber, \emph{Lectures at summer school on hadronic aspects of collider
  physics, {Z}uoz, {S}witzerland, {A}ugust 1994, hadronization},  1994.
\newblock hep-ph/9411384.

\bibitem{Dokshitzer_1995}
Y.~Dokshitzer and B.~Webber, \emph{Calculation of power corrections to hadronic
  event shapes},
  \href{https://doi.org/10.1016/0370-2693(95)00548-y}{\emph{Physics Letters B}
  {\bfseries 352} (1995) 451}.

\bibitem{Dokshitzer_1998}
Y.~Dokshitzer, A.~Lucenti, G.~Marchesini and G.~Salam, \emph{Universality of
  corrections to jet-shape observables rescued},
  \href{https://doi.org/10.1016/s0550-3213(97)00650-0}{\emph{Nuclear Physics B}
  {\bfseries 511} (1998) 396}.

\bibitem{Dokshitzer_1998_2}
Y.~L. Dokshitzer, A.~Lucenti, G.~Marchesini and G.~P. Salam, \emph{On the
  universality of the milan factor for 1/{Q} power corrections to jet shapes},
  \href{https://doi.org/10.1088/1126-6708/1998/05/003}{\emph{Journal of High
  Energy Physics} {\bfseries 1998} (1998) 003}.

\bibitem{Abbate:2012jh}
R.~Abbate, M.~Fickinger, A.~H. Hoang, V.~Mateu and I.~W. Stewart,
  \emph{{Precision Thrust Cumulant Moments at $N^3$LL}},
  \href{https://doi.org/10.1103/PhysRevD.86.094002}{\emph{Phys. Rev. D}
  {\bfseries 86} (2012) 094002}
  [\href{https://arxiv.org/abs/1204.5746}{{\ttfamily 1204.5746}}].

\bibitem{Hoang:2015hka}
A.~H. Hoang, D.~W. Kolodrubetz, V.~Mateu and I.~W. Stewart, \emph{{Precise
  determination of $\alpha_s$ from the $C$-parameter distribution}},
  \href{https://doi.org/10.1103/PhysRevD.91.094018}{\emph{Phys. Rev.}
  {\bfseries D91} (2015) 094018}
  [\href{https://arxiv.org/abs/1501.04111}{{\ttfamily 1501.04111}}].

\bibitem{Kardos:2018kqj}
A.~Kardos, S.~Kluth, G.~Somogyi, Z.~Tulip\'ant and A.~Verbytskyi,
  \emph{{Precise determination of $\alpha _{S}(M_Z)$ from a global fit of
  energy\textendash{}energy correlation to NNLO+NNLL predictions}},
  \href{https://doi.org/10.1140/epjc/s10052-018-5963-1}{\emph{Eur. Phys. J. C}
  {\bfseries 78} (2018) 498}
  [\href{https://arxiv.org/abs/1804.09146}{{\ttfamily 1804.09146}}].

\bibitem{Nason:2023asn}
P.~Nason and G.~Zanderighi, \emph{{Fits of $\alpha_s$ using power corrections
  in the three-jet region}},
  \href{https://arxiv.org/abs/2301.03607}{{\ttfamily 2301.03607}}.

\bibitem{Caola_2022}
F.~Caola, S.~F. Ravasio, G.~Limatola, K.~Melnikov and P.~Nason, \emph{On linear
  power corrections in certain collider observables},
  \href{https://doi.org/10.1007/jhep01(2022)093}{\emph{Journal of High Energy
  Physics} {\bfseries 2022} (2022) }.

\bibitem{Beneke:1998ui}
M.~Beneke, \emph{{Renormalons}},
  \href{https://doi.org/10.1016/S0370-1573(98)00130-6}{\emph{Phys. Rept.}
  {\bfseries 317} (1999) 1}
  [\href{https://arxiv.org/abs/hep-ph/9807443}{{\ttfamily hep-ph/9807443}}].

\bibitem{Dokshitzer_1996}
Y.~Dokshitzer, G.~Marchesini and B.~Webber, \emph{Dispersive approach to
  power-behaved contributions in {QCD} hard processes},
  \href{https://doi.org/10.1016/0550-3213(96)00155-1}{\emph{Nuclear Physics B}
  {\bfseries 469} (1996) 93}.

\bibitem{Nason_1995}
P.~Nason and M.~H. Seymour, \emph{Infrared renormalons and power suppressed
  effects in e+e- jet events},
  \href{https://doi.org/10.1016/0550-3213(95)00461-z}{\emph{Nuclear Physics B}
  {\bfseries 454} (1995) 291}.

\bibitem{Beneke:2000kc}
M.~Beneke and V.~M. Braun, \emph{{Renormalons and power corrections}},
  \href{https://arxiv.org/abs/hep-ph/0010208}{{\ttfamily hep-ph/0010208}}.

\bibitem{Schindler:2023cww}
S.~T. Schindler, I.~W. Stewart and Z.~Sun, \emph{{Renormalons in the
  energy-energy correlator}},
  \href{https://doi.org/10.1007/JHEP10(2023)187}{\emph{JHEP} {\bfseries 10}
  (2023) 187} [\href{https://arxiv.org/abs/2305.19311}{{\ttfamily
  2305.19311}}].

\bibitem{Dasgupta:2009tm}
M.~Dasgupta and Y.~Delenda, \emph{{On the universality of hadronisation
  corrections to QCD jets}},
  \href{https://doi.org/10.1088/1126-6708/2009/07/004}{\emph{JHEP} {\bfseries
  07} (2009) 004} [\href{https://arxiv.org/abs/0903.2187}{{\ttfamily
  0903.2187}}].

\bibitem{Dasgupta_2008}
M.~Dasgupta, L.~Magnea and G.~P. Salam, \emph{Non-perturbative {QCD} effects in
  jets at hadron colliders},
  \href{https://doi.org/10.1088/1126-6708/2008/02/055}{\emph{Journal of High
  Energy Physics} {\bfseries 2008} (2008) 055}.

\bibitem{Salam_2010}
G.~P. Salam, \emph{Towards jetography},
  \href{https://doi.org/10.1140/epjc/s10052-010-1314-6}{\emph{The European
  Physical Journal C} {\bfseries 67} (2010) 637}.

\bibitem{Dasgupta_2013}
M.~Dasgupta, A.~Fregoso, S.~Marzani and G.~P. Salam, \emph{Towards an
  understanding of jet substructure},
  \href{https://doi.org/10.1007/jhep09(2013)029}{\emph{Journal of High Energy
  Physics} {\bfseries 2013} (2013) }.

\bibitem{Marzani:2019hun}
S.~Marzani, G.~Soyez and M.~Spannowsky, \emph{{Looking inside jets: an
  introduction to jet substructure and boosted-object phenomenology}},
  vol.~958. Springer, 2019,
  \href{https://doi.org/10.1007/978-3-030-15709-8}{10.1007/978-3-030-15709-8},
  [\href{https://arxiv.org/abs/1901.10342}{{\ttfamily 1901.10342}}].

\bibitem{Bigi:1994em}
I.~I.~Y. Bigi, M.~A. Shifman, N.~G. Uraltsev and A.~I. Vainshtein, \emph{{The
  Pole mass of the heavy quark. Perturbation theory and beyond}},
  \href{https://doi.org/10.1103/PhysRevD.50.2234}{\emph{Phys. Rev. D}
  {\bfseries 50} (1994) 2234}
  [\href{https://arxiv.org/abs/hep-ph/9402360}{{\ttfamily hep-ph/9402360}}].

\bibitem{Webber:1994cp}
B.~R. Webber, \emph{{Estimation of power corrections to hadronic event
  shapes}}, \href{https://doi.org/10.1016/0370-2693(94)91147-9}{\emph{Phys.
  Lett. B} {\bfseries 339} (1994) 148}
  [\href{https://arxiv.org/abs/hep-ph/9408222}{{\ttfamily hep-ph/9408222}}].

\bibitem{Akhoury:1995sp}
R.~Akhoury and V.~I. Zakharov, \emph{{On the universality of the leading, 1/Q
  power corrections in QCD}},
  \href{https://doi.org/10.1016/0370-2693(95)00866-J}{\emph{Phys. Lett. B}
  {\bfseries 357} (1995) 646}
  [\href{https://arxiv.org/abs/hep-ph/9504248}{{\ttfamily hep-ph/9504248}}].

\bibitem{Dasgupta_1998}
M.~Dasgupta and B.~Webber, \emph{Power corrections to event shapes in deep
  inelastic scattering}, \href{https://doi.org/10.1007/s100520050103}{\emph{The
  European Physical Journal C} {\bfseries 1} (1998) 539}.

\bibitem{Amati:1980ch}
D.~Amati, A.~Bassetto, M.~Ciafaloni, G.~Marchesini and G.~Veneziano, \emph{{A
  Treatment of Hard Processes Sensitive to the Infrared Structure of QCD}},
  \href{https://doi.org/10.1016/0550-3213(80)90012-7}{\emph{Nucl. Phys. B}
  {\bfseries 173} (1980) 429}.

\bibitem{Beneke:1995pq}
M.~Beneke and V.~M. Braun, \emph{{Power corrections and renormalons in
  Drell-Yan production}},
  \href{https://doi.org/10.1016/0550-3213(95)00439-Y}{\emph{Nucl. Phys. B}
  {\bfseries 454} (1995) 253}
  [\href{https://arxiv.org/abs/hep-ph/9506452}{{\ttfamily hep-ph/9506452}}].

\bibitem{Nason:1995np}
P.~Nason and M.~H. Seymour, \emph{{Infrared renormalons and power suppressed
  effects in e+ e- jet events}},
  \href{https://doi.org/10.1016/0550-3213(95)00461-Z}{\emph{Nucl. Phys. B}
  {\bfseries 454} (1995) 291}
  [\href{https://arxiv.org/abs/hep-ph/9506317}{{\ttfamily hep-ph/9506317}}].

\bibitem{Larkoski:2014wba}
A.~J. Larkoski, S.~Marzani, G.~Soyez and J.~Thaler, \emph{{Soft Drop}},
  \href{https://doi.org/10.1007/JHEP05(2014)146}{\emph{JHEP} {\bfseries 05}
  (2014) 146} [\href{https://arxiv.org/abs/1402.2657}{{\ttfamily 1402.2657}}].

\bibitem{Catani:1993hr}
S.~Catani, Y.~L. Dokshitzer, M.~H. Seymour and B.~R. Webber,
  \emph{{Longitudinally invariant $K_t$ clustering algorithms for hadron hadron
  collisions}}, \href{https://doi.org/10.1016/0550-3213(93)90166-M}{\emph{Nucl.
  Phys. B} {\bfseries 406} (1993) 187}.

\bibitem{Ellis:1993tq}
S.~D. Ellis and D.~E. Soper, \emph{{Successive combination jet algorithm for
  hadron collisions}},
  \href{https://doi.org/10.1103/PhysRevD.48.3160}{\emph{Phys. Rev. D}
  {\bfseries 48} (1993) 3160}
  [\href{https://arxiv.org/abs/hep-ph/9305266}{{\ttfamily hep-ph/9305266}}].

\bibitem{Cacciari:2008gp}
M.~Cacciari, G.~P. Salam and G.~Soyez, \emph{{The anti-$k_t$ jet clustering
  algorithm}}, \href{https://doi.org/10.1088/1126-6708/2008/04/063}{\emph{JHEP}
  {\bfseries 04} (2008) 063} [\href{https://arxiv.org/abs/0802.1189}{{\ttfamily
  0802.1189}}].

\bibitem{Dokshitzer:1997in}
Y.~L. Dokshitzer, G.~Leder, S.~Moretti and B.~Webber, \emph{{Better jet
  clustering algorithms}},
  \href{https://doi.org/10.1088/1126-6708/1997/08/001}{\emph{JHEP} {\bfseries
  08} (1997) 001} [\href{https://arxiv.org/abs/hep-ph/9707323}{{\ttfamily
  hep-ph/9707323}}].

\bibitem{Wobisch:1998wt}
M.~Wobisch and T.~Wengler, \emph{{Hadronization corrections to jet
  cross-sections in deep inelastic scattering}},  in \emph{{Workshop on Monte
  Carlo Generators for HERA Physics (Plenary Starting Meeting)}}, pp.~270--279,
  4, 1998, \href{https://arxiv.org/abs/hep-ph/9907280}{{\ttfamily
  hep-ph/9907280}}.

\bibitem{Dasgupta:2013ihk}
M.~Dasgupta, A.~Fregoso, S.~Marzani and G.~P. Salam, \emph{{Towards an
  understanding of jet substructure}}, {\emph{JHEP} {\bfseries 09} (2013) 029}
  [\href{https://arxiv.org/abs/1307.0007}{{\ttfamily 1307.0007}}].

\bibitem{Hoang:2019ceu}
A.~H. Hoang, S.~Mantry, A.~Pathak and I.~W. Stewart, \emph{{Nonperturbative
  Corrections to Soft Drop Jet Mass}},
  \href{https://arxiv.org/abs/1906.11843}{{\ttfamily 1906.11843}}.

\bibitem{Pathak:2020iue}
A.~Pathak, I.~W. Stewart, V.~Vaidya and L.~Zoppi, \emph{{EFT for Soft Drop
  Double Differential Cross Section}},
  \href{https://doi.org/10.1007/JHEP04(2021)032}{\emph{JHEP} {\bfseries 04}
  (2021) 032} [\href{https://arxiv.org/abs/2012.15568}{{\ttfamily
  2012.15568}}].

\bibitem{Pathak:2023sgi}
A.~Pathak, \emph{{The Catchment Area of Groomed Jets at NNLL}},
  \href{https://arxiv.org/abs/2301.05714}{{\ttfamily 2301.05714}}.

\bibitem{Ferdinand:2023vaf}
A.~Ferdinand, K.~Lee and A.~Pathak, \emph{{Field-Theoretic Analysis of
  Hadronization Using Soft Drop Jet Mass}},
  \href{https://arxiv.org/abs/2301.03605}{{\ttfamily 2301.03605}}.

\bibitem{Catani_1999}
S.~Catani and M.~Grazzini, \emph{Collinear factorization and splitting
  functions for next-to-next-to-leading order {QCD} calculations},
  \href{https://doi.org/10.1016/s0370-2693(98)01513-5}{\emph{Physics Letters B}
  {\bfseries 446} (1999) 143}.

\end{thebibliography}\endgroup
\end{document}